\documentclass[twocolumn,prb]{revtex4}%
\usepackage[dvipdfmx]{graphicx}%
\usepackage{amsmath}%
\usepackage{color}
\usepackage{amsfonts}
\usepackage{amssymb}
\usepackage{hhline}
\usepackage{braket}
\usepackage{bm}
\usepackage{mathrsfs}

\begin{document}
\title{Fragment-orbital-dependent spin fluctuations in the single-component molecular conductor [Ni(dmdt)$_2$]}
\author{
Taiki Kawamura and Akito Kobayashi}
\address{
Department of Physics, Nagoya University, Nagoya, Aichi 464-8602, Japan
}
\begin{abstract}
Motivated by recent nuclear magnetic resonance experiments, we calculated the spin susceptibility, Knight shift, and spin-lattice relaxation rate ($1/T_{1}T$) of the single-component molecular conductor [Ni(dmdt)$_2$] using the random phase approximation in a multi-orbital Hubbard model describing the Dirac nodal line electronic system in this compound.
This Hubbard model is composed of three fragment orbitals and on-site repulsive interactions obtained using ab initio many-body perturbation theory calculations.
We found fragment-orbital-dependent spin fluctuations with the momentum $\textbf{q}$=$\textbf{0}$ and an incommensurate value of the wavenumber $\textbf{q}$=$\textbf{Q}$ at which a diagonal element of the spin susceptibility is maximum.
The $\textbf{q}$=$\textbf{0}$ and $\textbf{Q}$ responses become dominant at low and high temperatures, respectively, with the Fermi-pocket energy scale as the boundary.
We show that $1/T_{1}T$ decreases with decreasing temperature but starts to increase at low temperature owing to the $\textbf{q}$=$\textbf{0}$ spin fluctuations, while the Knight shift keeps monotonically decreasing.
These properties are due to the intra-molecular antiferromagnetic fluctuations caused by the characteristic wave functions of this Dirac nodal line system, which is described by an $n$-band ($n\geq 3$) model. We show that the fragment orbitals play important roles in the magnetic properties of [Ni(dmdt)$_2$].
\end{abstract}

\sloppy

\maketitle

\section{introduction}
Dirac electron systems in solids are of interest to many researchers because of not only their quantum transport phenomena \cite{T.Ando1998,V.P.Gusynin2006,S.Murakami2004,D.Hsieh2008} and large diamagnetism, \cite{H.Fukuyama1970,H.Fukuyama2007} but also their unusual effects induced by the Coulomb interaction.\cite{A.A.Abrikosov1971,J.Gonzalez1994,J.Gonzalez1999,V.N.Kotov2012,T.O.Wehling2014,W.Witczak-Krempa2014}

Dirac electron systems in molecular conductors, such as $\alpha$-(BEDT-TTF)$_2$I$_3$, provide suitable platforms for studying the effect of interaction because the electron
hopping integrals between neighboring molecules are smaller than the on-site repulsive interactions
reflecting the weak inter-molecular coupling.
\cite{K.Kajita1992,N.Tajima2000,A.Kobayashi2004,S.Katayama2006,
A.Kobayashi2007,M.O.Goerbig2008,K.Kajita2014}
At high pressure, $\alpha$-(BEDT-TTF)$_2$I$_3$ is a massless Dirac electron system.
However, at low pressure, a charge-ordered state appears presumably due to nearest-neighbor Coulomb repulsions,
\cite{H.Seo2000,T.Takahashi2003,T.Takiuchi2007}
where anomalous spin--charge separation on spin gaps \cite{K.Ishibashi2016,Y.Katayama2016} and transport phenomena occur.\cite{R.Beyer2016,D.Lu2016,D.Ohki2019}
In addition, the long-range Coulomb interaction reshapes the Dirac cone because of a logarithmic velocity renormalization, which induces an anomalous magnetic response.\cite{M.Hirata2016,M.Hirata2017,A.Kobayashi2013} Moreover, ferrimagnetism and spin-triplet excitonic fluctuations are observed.\cite{D.Ohki2020,M.HirataRPP}

The Dirac electron system in $\alpha$-(BEDT-TTF)$_2$I$_3$ is two-dimensional \cite
{K.Kajita2014} because it is a layered molecular conductor and the hopping of electrons from one conducting layer to the neighboring one over the insulating anion layer is incoherent.
By contrast, if the electron hopping perpendicular to the main conducting layer were coherent, the Dirac point would be connected and draw lines (rings) in the three-dimensional momentum space, which are called the Dirac nodal lines (rings). \cite{A.Burkov2011,C.-K.Chiu2014,C.Fang2015,Z.Gao2016}

Such kinds of Dirac nodal line (ring) systems have indeed been found in graphite,
\cite{P.R.Wallace1947} transition-metal monophosphates,\cite{H.Weng2015} Cu$_{3}$N,\cite{Y.Kim2015} antiperovskites,\cite{R.Yu2015} perovskite iridates,\cite{J.-M.Carter2012} and hexagonal pnictides with the composition CaAgX (X = P, As),\cite{A.Yamakage2016} as well as in the single-component molecular conductors [Pd(dddt)$_2$], \cite{R.Kato2017_A,R.Kato2017_B,Y.Suzumura2017_A,Y.Suzumura2017_B,Y.Suzumura2018_A,Y.Suzumura2018_B,T.Tsumuraya2018,Y.Suzumura2019} [Pt(dmdt)$_2$],\cite{B.Zhou2019,R.Kato2020,T.Kawamura2020,T.Kawamura2021,A.Kobayashi2021} and [Ni(dmdt)$_2$].\cite{T.Kawamura2021, A.Kobayashi2021}

The Dirac nodal line (ring) systems exhibit not only the properties in common with two-dimensional Dirac electron systems, {\it e.g.}, the in-plane conductivity\cite{B.Zhou2019}, but also the characteristic electronic properties such as non-dispersive Landau levels,\cite{J.-W.Rhim2015} Kondo effect,\cite{A.K.Mitchell2015} quasi-topological electromagnetic responses,\cite{S.T.Ramamurthy2017} and edge magnetism\cite{,T.Kawamura2021} because of the three-dimensionality.
However, the electron correlation effects on the Fermi surface in the Dirac nodal line systems have not yet been elucidated.

\begin{figure}[htpb]
\begin{center}
\includegraphics[width=80mm]{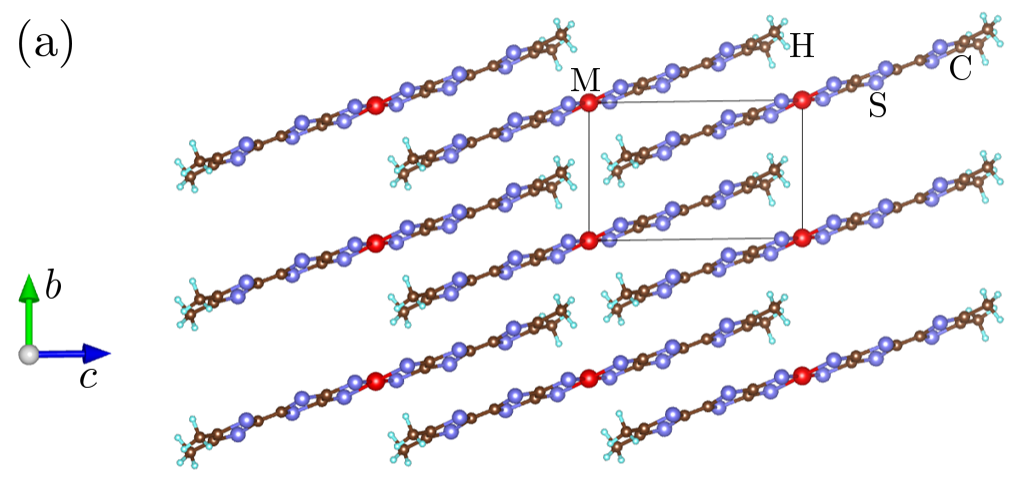}
\end{center}
\begin{center}
\includegraphics[width=50mm]{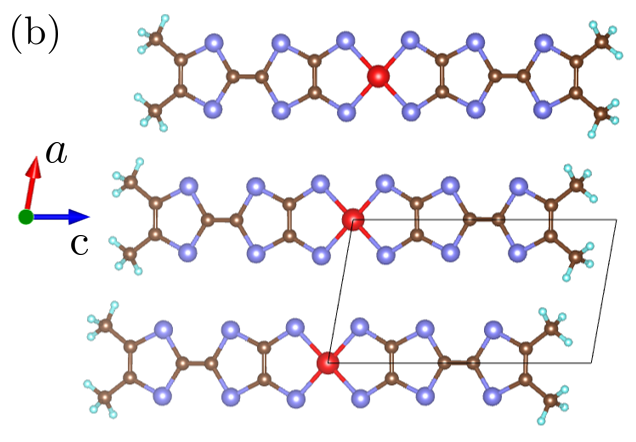}
\end{center}
\caption{ (a) Crystal structure in the $b$--$c$ plane of [M(dmdt)$_2$]. This material consists of M(dmdt)$_2$ molecules.
(b) Crystal structure along to the $b$-axis.
The red, blue, brown, and cyan balls represent Ni, S, C, and H atoms, respectively.
The black frames represent the unit cell.
}
\label{Crystal}
\end{figure}

The prime focus of this study is such Dirac nodal line system in [M(dmdt)2] (M = Pt, Ni), which is a single-component molecular conductor that consists of the M(dmdt)$_2$ molecules, where the bracket [$\cdots$] stand for a crystal.
This material is a triclinic system, as shown in Fig. \ref{Crystal}, and has space-inversion symmetry. One unit cell contains one M(dmdt)$_2$ molecule.
In previous studies, the electronic properties of [M(dmdt)$_2$] were studied using density functional theory (DFT), and tight-binding models were constructed on the basis of the extended H\"uckel method and DFT.\cite{B.Zhou2019, R.Kato2020, T.Kawamura2020, T.Kawamura2021} These investigations showed that [M(dmdt)$_2$] is a Dirac nodal line system. Furthermore, electronic resistivity measurements using conventional four-probe methods were performed and showed that the resistivity of [M(dmdt)$_2$] hardly depends on the temperature ($T$), which is consistent with the property of the Dirac electron system.\cite{B.Zhou2019} That is the universal conductivity.\cite{N.H.Shon1998}
In addition, we previously suggested that the nesting between the Fermi arcs localized at the edge and the electronic correlation induce a helical spin density wave (SDW) at the edge.\cite{T.Kawamura2021}

Recently, the spin-lattice relaxation rate $1/T_{1}T$, probing the low-energy spin dynamics, and the Knight shift, scaling to the spin susceptibility, of [Ni(dmdt)$_2$] were observed in a $^{13}$C nuclear magnetic resonance ($^{13}$C-NMR) experiment.\cite{T.Sekine.Private} At high temperature, $1/T_{1}T$ decreases with cooling and is almost proportional to $T^{2}$. However, at low temperature, it starts to increase with decreasing temperature and exhibits a peak structure at approximately $30$ K.
Meanwhile, the Knight shift is almost proportional to $T$ because of the linear energy dispersion and does not increase. The mechanism of this anomalous temperature dependence of the spin fluctuations has not been elucidated.\\

In the present study, we theoretically investigate the electron correlation using the Fermi surface in [Ni(dmdt)$_2$] to elucidate the mechanism of this anomalous temperature dependence of the spin fluctuations.
We calculate the spin susceptibility, Knight shift, and $1/T_{1}T$ using the random phase approximation (RPA) in a three-orbital Hubbard model describing [Ni(dmdt)$_2$], which is obtained using ab initio many-body perturbation theory calculations.

The electronic state of a molecular conductor is described by the molecular orbitals, which are linear combinations of the atomic orbitals in a molecule. The molecular orbital that has the highest energy and is fully
occupied by electrons is called the highest occupied molecular orbital (HOMO), whereas the one having the lowest energy with no electrons is called the lowest unoccupied molecular orbital (LUMO). HOMO and LUMO are also called frontier orbitals.
The electronic states of single-component molecular conductors, {\it e.g.}, [M(tmdt)$_2$] (M = Ni, Au, Cu) and [M(dmdt)$_2$] (M = Pt, Ni), are described by multiple molecular orbitals localized in a part of the molecule.\cite{H.Seo2008, H.Seo2013, M.Tsuchiizu2012, T.Kawamura2020, T.Kawamura2021}
These molecular orbitals are the energy eigenstates obtained using first-principles calculations and are called ``fragment orbitals''.
The fragment orbitals are transformed into HOMO and LUMO by a high-symmetry unitary transform.

Based on the band parameters determined from first-principle calculations, Seo et al.
have constructed a Hubbard model of [M(tmdt)$_2$](M=Ni, Au, Cu), which is described by the fragment orbitals.\cite{H.Seo2008, H.Seo2013} The on-site repulsion acts between the same fragment orbitals that have spins of opposite signs, which is similar to the case of the present study. They have investigated the ordered state by calculating the electron density and spin density using mean-field approximation. By contrast, in this study, we will investigate the spin fluctuations by calculating the spin susceptibility on the basis of RPA. Furthermore, we show that the idea of the fragment orbitals is important for the physical properties of the Dirac nodal line systems in the single-component molecular conductor.
As the other previous study of the fragment orbitals, some charge-transfer complexes such as (TTM-TTP)I$_3$ are also modeled by fragment orbitals.\cite{M.Tsuchiizu2011, M.Tsuchiizu2012}

We found that the $\textbf{q}$=$\textbf{0}$ spin fluctuations are enhanced in two out of the three fragment orbitals, while an enhancement at an incommensurate wavenumber vector develops in the third orbital.
Detailed analysis showed that these $\textbf{q}$=$\textbf{0}$ spin fluctuations do not correspond to a simple ferromagnetic fluctuations; rather, they are linked to an intra-molecular antiferromagnetic fluctuations. This implies that the spins of the fragment orbitals within the same molecule are inversely correlated.
Further, $\textbf{q}$=$\textbf{0}$ implies a direct correlation of the spins between molecules.
Using RPA, we determined that the $1/T_{1}T$ starts to increase at a low temperature by the $\textbf{q}$=$\textbf{0}$ spin fluctuations.
By contrast, the Knight shift does not increase upon cooling because of the intra-molecular antiferromagnetic fluctuations.

At high temperature, an incommensurate spin fluctuations dominate the temperature
dependence of $1/T_{1}T$. These magnetic responses are associated with the geometry of the Fermi surface and the characteristic wave functions of the $n$-orbital ($n\geq 3$) Dirac nodal line system. Thus, it is expected that other Dirac nodal line systems described by multiple-orbital models may have similar magnetic properties.

The remainder of this paper is organized as follows. In Section I\hspace{-.1em}I, we introduce the spin susceptibility based on RPA and formulate $1/T_{1}T$ and the Knight shift. In Section I\hspace{-.1em}I\hspace{-.1em}I, we calculate the band structure, spin susceptibility, Knight shift, $1/T_{1}T$, and so on in the absence of interaction. In Section I\hspace{-.1em}V, we calculate the Stoner factor, Knight shift, and $1/T_1T$ in the presence of interaction by applying RPA to a Hubbard model. Section V draws conclusions.

\section{Formulation}
We calculate the spin susceptibility, which incorporates the electron correlation effect within perturbation theory to investigate the enhancement of spin fluctuations in [Ni(dmdt)$_2$]. Furthermore, we calculate the Knight shift and the spin-lattice relaxation rate $1/T_{1}T$, which are the physical quantities observed using NMR. We apply RPA to the Hubbard model to calculate the spin susceptibility in [Ni(dmdt)$_2$]. Although calculations incorporating the self-energy, {\it e.g.}, FLEX, are better approximations than RPA, RPA is more suitable for investigating spin fluctuations because the self-energy suppresses spin susceptibility.

\begin{figure}[htpb]
\begin{center}
\includegraphics[width=80mm]{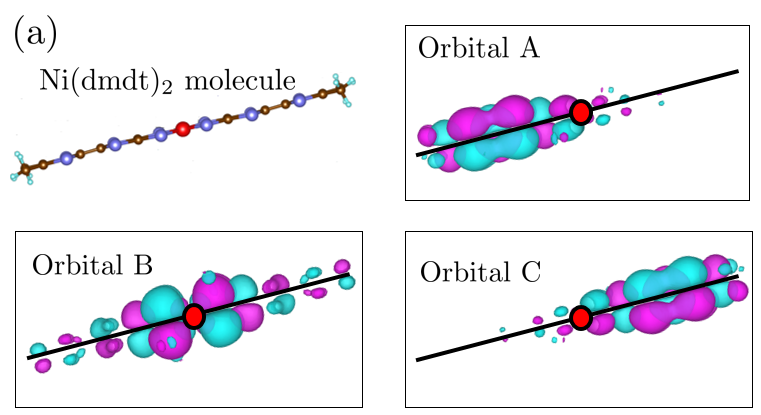}
\end{center}
\begin{center}
\includegraphics[width=80mm]{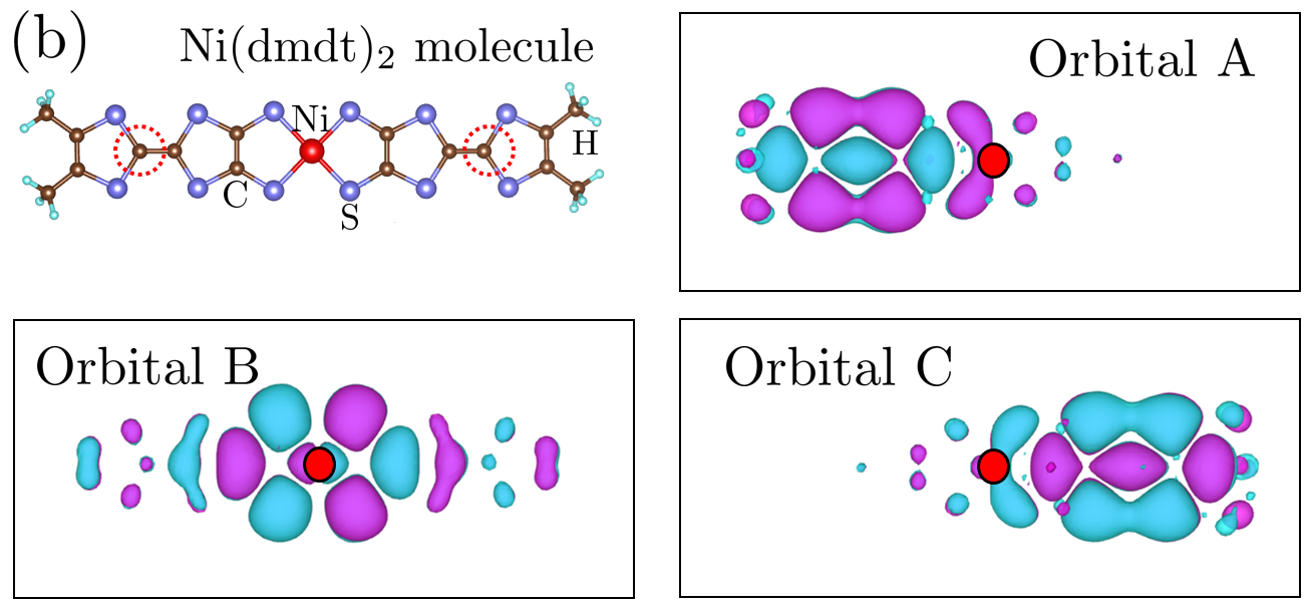}
\end{center}
\caption{ Schematic of a Ni(dmdt)$_2$ molecule and the fragment orbitals A, B, and C.
The red, blue, brown, and cyan balls represent Ni, S, C, and H atoms, respectively.
The red dots in orbitals A, B, and C indicate the location of the Ni atom as a guide to the eye. (a) Side view of the molecule. The black solid line represents the Ni(dmdt)$_2$ molecule. (b) Vertical-axis view of the molecule. The red dashed circles indicate C atoms, which are replaced with $^{13}$C in the $^{13}$C-NMR experiment.\cite{T.Sekine.Private} These figures are plotted by VESTA. Different colors represent different signs of the wave functions. The cut-off of the normalized wave functions is 0.01.
}
\label{fragment}
\end{figure}

The Hubbard Hamiltonian that we employ is given by
\begin{equation}
\label{H_R}
H=\sum_{\left< i,\alpha;j,\beta \right>,\sigma}t_{i,\alpha;j,\beta}c^{\dagger}_{i,\alpha,\sigma}c_{j,\beta,\sigma}+\sum_{i,\alpha}U_{\alpha}n_{i,\alpha,\uparrow}n_{i,\alpha,\downarrow},
\end{equation}
where $i$ and $j$ are the unit-cell indices, and $\sigma$ is the spin index.
Here, $t_{i,\alpha;j,\beta}$ is a transfer integral defined between the orbital $\alpha$ in the unit cell $i$ and the orbital $\beta$ in the
cell $j$, and $U_{\alpha}$ represents the on-site repulsive interaction on the orbital $\alpha$, with $\alpha$ and $\beta$ standing for one of the three fragment orbitals in the unit cell (A, B, and C in Fig. \ref{fragment}).
The indices $\alpha$ and $\beta$ represent the three fragment orbitals A, B, and C.
$\sum_{\left<\cdots \right>}$ represents a summation that runs only for the hoppings that have a large energy scale than the cutoff (set to be $0.010$ eV in this study).

The electronic states of [Ni(dmdt)$_2$] near the Fermi energy $E_{F}$ are described by three fragment-decomposed Wannier orbitals(fragment orbitals) dubbed orbitals A, B, and C as illustrated in Fig. \ref{fragment}. They are obtained using Wannier fitting to three isolated energy bands near $E_{F}$, which were previously obtained using first-principles calculations. \cite{T.Kawamura2021}

The Wannier fitting and first-principles calculations were performed using the programs \textsc{respack}\cite{K.Nakamura2020} and \textsc{Quantum} \textsc{Espresso}\cite{P.Giannozzi2017}, respectively.\cite{T.Kawamura2021} \textsc{respack} was also used for calculating the Coulomb interaction and other factors. Further, \textsc{Quantum} \textsc{Espresso} was used for first-principles calculations based on the pseudopotential method.
Figure \ref{fragment}(a) and (b) show a side and vertical-axis view, respectively, of the molecule.
The orbital B in Fig. \ref{fragment}(a) may look like a $d$ orbital, but is a $p$ orbital of
the S atoms, as is evident from Fig. \ref{fragment}(b).
The $d$ orbital of Ni is localized near the Ni atom, and its contribution to the orbital B is small. The orbitals A and C have $p$-orbital-like shapes and are equivalent because of space-inversion symmetry.
In the present study, we assume that these three fragment orbitals sit in the same molecule and in the same unit cell.

\begin{figure}[htpb]
\begin{center}
\includegraphics[width=75mm]{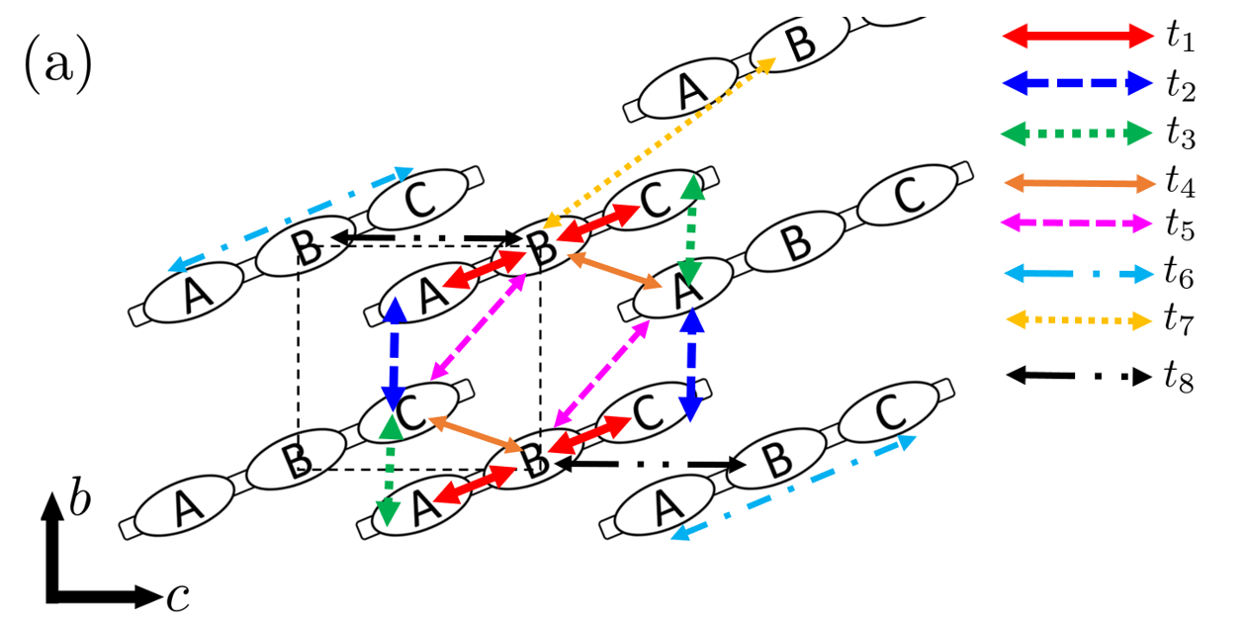}
\end{center}
\begin{center}
\includegraphics[width=75mm]{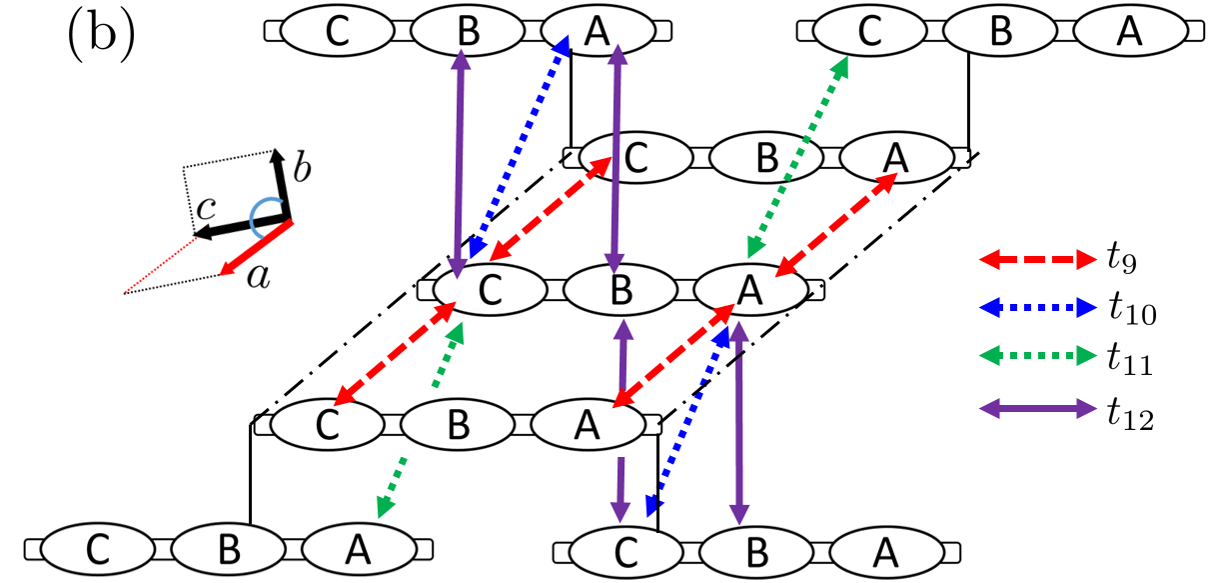}
\end{center}
\caption{
Hopping networks between fragment orbitals in tight-binding model of [Ni(dmdt)2]. (a) Schematic
illustration of the 2D network of the transfer integrals (shown by double--headed arrows) in the
crystalline $bc$--plane. The dashed square stands for the unit cell. (b) Schematic illustration of the 3D
hopping network including the transfer integrals along the $a$--direction. The black chain lines and vertical bold lines in (b) are guides to the eyes: The former lines are parallel to the $a$--direction, while the latter lines connect the molecules in the same $bc$plane.
}
\label{network}
\end{figure}

By performing a Fourier transform, the Hamiltonian (Eq. \ref{H_R}) is rewritten as
\begin{equation}
\label{H_K}
\begin{split}
H&=\sum_{\textbf{k},\alpha,\beta,\sigma}H^{0}_{\alpha\beta,\sigma}(\textbf{k})c^{\dagger}_{\textbf{k},\alpha,\sigma}c_{,\textbf{k},\beta,\sigma} \\
&+\frac{1}{N_L}\sum_{\textbf{k},\textbf{k}',\textbf{q},\alpha}U_{\alpha\alpha}c^{\dagger}_{\textbf{k}+\textbf{q},\alpha,\uparrow}c^{\dagger}_{\textbf{k}'-\textbf{q},\alpha,\downarrow}c_{\textbf{k}',\alpha,\downarrow}c_{\textbf{k},\alpha,\uparrow},
\end{split}
\end{equation}
where $\textbf{k}$, $\textbf{k}'$, and $\textbf{q}$ are the wavenumber vectors.
$N_L$ is the number of the unit cells in the system.
The first term corresponds to the unperturbed Hamiltonian, and the second term is treated as a perturbed Hamiltonian.
Here, $H^{0}_{\alpha\beta,\sigma}(\textbf{k})$ is defined as
\begin{equation}
\label{transfer_definition}
H^{0}_{\alpha\beta,\sigma}(\textbf{k})=\sum_{\left< {\boldsymbol \delta} \right>}t_{\alpha\beta,\sigma, {\boldsymbol \delta}}e^{i\textbf{k}\cdot {\boldsymbol \delta}},
\end{equation}
where $ {\boldsymbol \delta}$ is a lattice vector connecting the neighbor unit cell. Further, $t_{\alpha\beta,\sigma, {\boldsymbol \delta}}$ is the transfer integral between the fragment orbitals $\alpha$ and $\beta$, which are separated by the lattice vector $ {\boldsymbol\delta}$ and have the spin $\sigma$. We allot $t_{\alpha\beta,\sigma, {\boldsymbol \delta}}$ to the transfer integrals $t_1$, $t_2$, ..., $t_{12}$ and the site potential $\Delta$ in Table \ref{transfer},
where $\Delta \equiv t_{BB,\sigma,\textbf{0}}-t_{AA,\sigma,\textbf{0}}$.
Note that these transfer integrals were obtained from the Wannier fitting.
And we omit the small hoppings whose sizes are less than a cutoff energy of $0.010$ eV to make the analysis simple (see Fig. \ref{network} for the schematic illustration of such a hopping network).
Note that $t_{1}$ connects the nearest-neighbor fragment orbitals within a molecule, while $t_{2}$, $t_{3}$, ..., $t_{8}$ connect those between molecules in the crystalline $b$--$c$ plane, which corresponds to the $k_b$--$k_c$ plane in momentum space, where Dirac cones exist. We point out that $t_1$, $t_2$, and $t_3$ are the three essential transfer integrals that are needed to create the Dirac cones, while $t_9$ creates Fermi surfaces, and $t_{10}$, $t_{11}$, and $t_{12}$ wind the Dirac nodal lines.

\begin{table}[htb]
\begin{tabular}{|c|} \hline
~~~~~~~~transfer integrals (eV) ~~~~~~~\\ \hline
$t_{1}$~~~~~~~-0.2372\\
$t_{2}$~~~~~~~-0.1840\\
$t_{3}$~~~~~~~-0.2080\\
$t_{4}$~~~~~~~0.0302\\
$t_{5}$~~~~~~~0.0326\\
$t_{6}$~~~~~~~-0.0389\\
$t_{7}$~~~~~~~0.0103\\
$t_{8}$~~~~~~~-0.0144\\
$t_{9}$~~~~~~~-0.0140\\
$t_{10}$~~~~~~-0.0541\\
$t_{11}$~~~~~~-0.0534\\
$t_{12}$~~~~~~0.0116\\ \hline\hline
~~~~~~~~~site potential (eV)~~~~~~~ \\ \hline
$\Delta$~~~~~~~0.0429\\ \hline
\end{tabular}
\caption{Transfer integrals and site potential of [Ni(dmdt)$_2$].}
\label{transfer}
\end{table}
Previously, we calculated some quantities of [M(dmdt)$_2$] (M = Pt, Ni) considering the spin--orbit interaction (SOI) as a parameter. There, we found that the SOI can reduce the Fermi surface in the bulk and induce helical edge modes.\cite{T.Kawamura2020, T.Kawamura2021}
However, because the energy scale of SOI in this material ($\sim 0.0016$ eV) seems to be considerably smaller than the energy scale of the Fermi surface ($\sim 0.01$ eV),\cite{T.Kawamura2021} in reality SOI should not be large enough to significantly reduce the size of the Fermi surface.
Therefore, we will omit the influence of SOI on the Knight shift and $1/T_{1}T$.

From the definition in Eq. \ref{transfer_definition}, $H^{0}_{\alpha\beta,\sigma}(\textbf{k})$ are expressed by the following equations.
\begin{equation}
\label{H_element}
\begin{split}
H^{0}_{AA,\sigma}(\textbf{k})&=2t_{9}\cos{k_{a}}, \\
H^{0}_{AB,\sigma}(\textbf{k})&=t_{12}e^{i(-k_{a}+k_{b}+k_{c})}+t_{1}\\
&+t_{5}e^{i(k_{b}+k_{c})}+t_{4}e^{ik_{c}},\\
H^{0}_{AC,\sigma}(\textbf{k})&=t_{10}e^{i(-k_{a}+k_{b}+k_{c})}+t_{11}e^{i(k_{a}+k_{c})}+t_{6}\\
&+t_{3}e^{ik_{c}}+t_{2}e^{i(k_{b}+k_{c})},\\
H^{0}_{BB,\sigma}(\textbf{k})&= \Delta+2t_{7}\cos{(k_{b}+k_{c})}+2t_{8}\cos{k_{c}},\\
H^{0}_{BC,\sigma}(\textbf{k})&=t_{12}e^{i(-k_{a}+k_{b}+k_{c})}+t_{1}\\
&+t_{5}e^{i(k_{b}+k_{c})}+t_{4}e^{ik_{c}}, \\
H^{0}_{CC,\sigma}(\textbf{k})&=2t_{9}\cos{k_{a}}.
\end{split}
\end{equation}
Here, for simplicity, we set all the lattice constants to be unity.
Two of the three bands for which we performed Wannier fitting are occupied.\cite{T.Kawamura2020} Thus, the tight-binding model obtained using the Wannier fitting is also $2/3$ filling.
The unperturbed Hamiltonian $\hat{H}^{0}_{\sigma}(\textbf{k})$ satisfies the eigenvalue equation
\begin{equation}
\label{eigen_Eq}
\hat{H}^{0}_{\sigma}(\textbf{k})\ket{{\textbf{k},n,\sigma}}=E_{n,\sigma}(\textbf{k})\ket{{\textbf{k},n,\sigma}},
\end{equation}
\begin{equation}
\label{eigen_vec}
\ket{{\textbf{k},n,\sigma}}=
\left(
\begin{array}{c}
d_{A,n,\sigma}(\textbf{k}) \\
d_{B,n,\sigma}(\textbf{k}) \\
d_{C,n,\sigma}(\textbf{k})
\end{array}
\right).
\end{equation}
where $E_{n,\sigma}(\textbf{k})$ is the eigenvalue and $\ket{{\textbf{k},n,\sigma}}$ is the eigen vector; $n$ is the band index; and $d_{\alpha,n,\sigma}(\textbf{k})$ denotes the wave functions.
$\hat{H}^{0}_{\sigma}(\textbf{k})$ in Eq. \ref{eigen_Eq} consists of the matrix elements $H^{0}_{\alpha\beta,\sigma}(\textbf{k})$ in Eqs. \ref{H_K} and \ref{H_element}.
Reflecting the $2/3$-filling, the chemical potential $\mu$ is determined by
\begin{equation}
\label{standard}
\frac{1}{N_{L}}\sum_{\textbf{k},n,\sigma}f_{\textbf{k},n,\sigma}=4,
\end{equation}
\begin{equation}
\label{eigen_energy}
\epsilon_{\textbf{k},n,\sigma}\equiv E_{n,\sigma}(\textbf{k})-\mu,
\end{equation}
$f_{\textbf{k},n,\sigma}$=$1/\left[1+\exp{(\epsilon_{\textbf{k},n,\sigma}/T)}\right]$ is the Fermi distribution function, and we have $\mu$=$E_{F}$ at $T$=$0$.

As to the second term in the Hamiltonian Eq. \ref{H_K}, we introduce the on-site repulsive interaction $U_{\alpha\alpha}$ defined on the fragment orbital $\alpha$, which is defined as the
diagonal element of the interaction matrix as follows
\begin{eqnarray}
\label{OnsiteU}
\hat{U}&=& \left(
\begin{array}{ccc}
U_{AA} & 0 & 0 \\
0 & U_{BB} & 0 \\
0 & 0 & U_{CC} \\
\end{array}
\right) \\ \nonumber
&=& \left(
\begin{array}{ccc}
\lambda U & 0 & 0 \\
0 & U & 0 \\
0 & 0 & \lambda U \\
\end{array}
\right),
\end{eqnarray}
Here, because of the inversion symmetry, we have $U_{AA}$=$U_{CC}$, and we use the relative size of $U_{AA}$ to $U_{BB}$, $\lambda$=$U_{AA}/U_{BB}$, as a control parameter of this model.
According to \textsc{respack}, we found $U$=$6.72$ eV and $\lambda$=$0.79$ in the unscreened case, and $U$=$2.68$ eV and $\lambda$=$0.95$ in the screened case. In this study, we set $\lambda$=$0.95$ and use a value of $U$ less than $2.68$ eV because $\hat{U}$ tends to be overestimated in RPA.\\

\begin{figure}[htpb]
\begin{center}
\includegraphics[width=80mm]{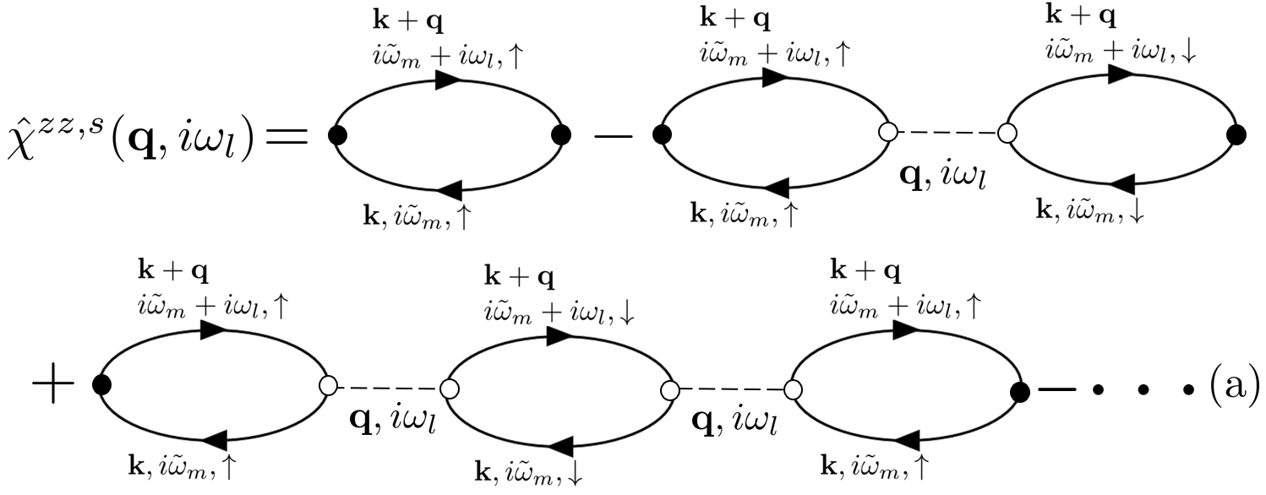}
\end{center}
\begin{center}
\includegraphics[width=75mm]{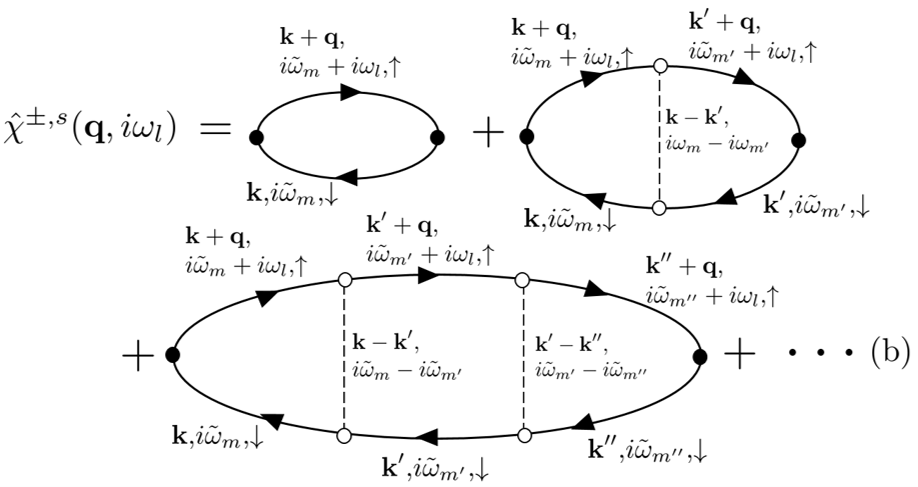}
\end{center}
\caption{
(A) Feynman diagram of $\hat{\chi}^{zz, s}(\textbf{q},i\omega_{l})$. (B) Feynman diagram of $\hat{\chi}^{\pm, s}(\textbf{q},i\omega_{l})$. The solid lines represent the non-interacting Green's functions, which describe the quasi-particles. The dashed lines represent the interactions.
The open circles are the vertexes connecting the non-interacting Green's functions and the interaction. The black dots represent the spin operators.
}
\label{diagram}
\end{figure}

The longitudinal and transverse spin susceptibilities are defined as follows:
\begin{eqnarray}
\label{chi_zz}
&\hat{\chi}^{zz}(\textbf{q},i\omega_{l}) \equiv \frac{1}{2}\int^{1/T}_{0}d\tau e^{i\omega_l \tau}\left<T_{\tau}\hat{S}^{z}_{\textbf{q}}(\tau)\hat{S}^{z}_{-\textbf{q}}(0)\right>, \\
&\hat{S}^{z}_{\textbf{q}}=\frac{1}{N_L}\sum_{\textbf{k}}\left(\hat{c}^{\dagger}_{\textbf{k}+\textbf{q},\uparrow}\hat{c}_{\textbf{k},\uparrow}-\hat{c}^{\dagger}_{\textbf{k}+\textbf{q},\downarrow}\hat{c}_{\textbf{k},\downarrow}\right),
\end{eqnarray}
\begin{eqnarray}
\label{chi_pm}
&\hat{\chi}^{\pm}(\textbf{q},i\omega_{l}) \equiv \int^{1/T}_{0}d\tau e^{i\omega_l \tau}\left<T_{\tau}\hat{S}^{+}_{\textbf{q}}(\tau)\hat{S}^{-}_{-\textbf{q}}(0)\right>,\\
&\hat{S}^{+}_{\textbf{q}}=\frac{1}{N_L}\sum_{\textbf{k}}\hat{c}^{\dagger}_{\textbf{k}+\textbf{q},\uparrow}\hat{c}_{\textbf{k},\downarrow},\\
&\hat{S}^{-}_{-\textbf{q}}=\frac{1}{N_L}\sum_{\textbf{k}}\hat{c}^{\dagger}_{\textbf{k},\downarrow}\hat{c}_{\textbf{k}+\textbf{q},\uparrow}.
\end{eqnarray}
Here, $i\omega_{l}$=$2li\pi T$ $(l\in\textbf{Z})$ is the Matsubara frequency and $\tau$ is the imaginary time.
$\hat{S}^{z}_{\textbf{q}}$, $\hat{S}^{+}_{\textbf{q}}$, and $\hat{S}^{-}_{\textbf{q}}$ are the spin operators. $\hat{S}^{z}_{\textbf{q}}(\tau)$, $\hat{S}^{+}_{\textbf{q}}(\tau)$, and $\hat{S}^{-}_{\textbf{q}}(\tau)$ are described in the Heisenberg picture.
The spin susceptibility is the proportionality coefficient of the magnetization to the infinitesimal magnetic field.
It represents the degree of the ``spin fluctuations'', because spins in the system sensitively respond to the infinitesimal magnetic field when spin susceptibility is large.
\\
By performing a perturbation expansion of Eqs. \ref{chi_zz} and \ref{chi_pm}, we obtain the non-interacting longitudinal spin susceptibility $\hat{\chi}^{zz,0}(\textbf{q},i\omega_{l})$ and non-interacting transverse spin susceptibility $\hat{\chi}^{\pm,0}(\textbf{q},i\omega_{l})$ as the zeroth-order perturbation terms.
In this study, SU(2) symmetry is protected. Therefore, we define $\hat{\chi}^{zz,0}(\textbf{q},i\omega_{l})$=$\hat{\chi}^{\pm,0}(\textbf{q},i\omega_{l})$$\equiv$ $\hat{\chi}^{0}(\textbf{q},i\omega_{l})$. Its matrix elements are written as
\begin{equation}
\label{chi0_formula}
\begin{split}
&\chi^{0}_{\alpha\beta}(\textbf{q},i\omega_{l}) \\
&=-\frac{T}{N_L}\sum_{\textbf{k},m}G^{0}_{\alpha\beta}(\textbf{k}+\textbf{q},i\tilde{\omega}_{m}+i\omega_{l})G^{0}_{\beta\alpha}(\textbf{k},i\tilde{\omega}_{m}),
\end{split}
\end{equation}
\begin{equation}
\label{Green}
\begin{split}
G^{0}_{\alpha\beta,\sigma}(\textbf{k},i\tilde{\omega}_{l})=\sum_{n}d_{\alpha,n,\sigma}(\textbf{k})d^{*}_{\beta,n,\sigma}(\textbf{k})\frac{1}{i\tilde{\omega}_{l}-\epsilon_{\textbf{k},n,\sigma}}.
\end{split}
\end{equation}
Eq. \ref{Green} expresses the matrix elements of the non-interacting Green's function. $i\tilde{\omega}_{l}$=$(2l+1)i\pi T$ is the Matsubara frequency of the fermions.
The spin index $\sigma$ is omitted in Eq. \ref{chi0_formula} because we have $\epsilon_{\textbf{k},n,\uparrow}$=$\epsilon_{\textbf{k},n,\downarrow}$ in Eq. \ref{Green}.
The longitudinal and transverse spin susceptibilities in RPA are represented by the Feynman diagrams shown in Fig. \ref{diagram}.
The first terms in the right-hand sides of diagrams (A) and (B) correspond to the terms $\hat{\chi}^{zz,0}(\textbf{q},i\omega_{l})$=$\hat{\chi}^{\pm,0}(\textbf{q},i\omega_{l})$$=$ $\hat{\chi}^{0}(\textbf{q},i\omega_{l})$ (Eq. \ref{chi0_formula}).
Because the interacting longitudinal and transverse spin susceptibilities are represented by summations of the series of $\hat{U}\hat{\chi}^{0}(\textbf{q},i\omega_{l})$ in RPA, they are written as
\begin{eqnarray}
\label{chiS_RPA}
\hat{\chi}^{zz,s}(\textbf{q},i\omega_{l})&=&\hat{\chi}^{\pm,s}(\textbf{q},i\omega_{l})\equiv\hat{\chi}^{s}(\textbf{q},i\omega_{l}) \\ \nonumber
&=&\hat{\chi}^{0}(\textbf{q},i\omega_{l})[\hat{I}-\hat{U}\hat{\chi}^{0}(\textbf{q},i\omega_{l})]^{-1} \\
, \nonumber
\end{eqnarray}
where $\hat{I}$ is the unit matrix.

Here we introduce the Stoner factor $\xi_s(\textbf{q})$ representing the degree of enhancement of the spin fluctuations.
The Stoner factor $\xi_{s}(\textbf{q})$ is defined as the maximum eigenvalue of $\hat{U}\hat{\chi}^{0}(\textbf{q},0)$.
The relation between $\xi_{s}(\textbf{q})$ and $\hat{\chi}^{s}(\textbf{q},0)$ in the three-orbital model is given by
\begin{eqnarray}
\label{xi_and_chiS}
\hat{\chi}^{s}(\textbf{q},0)=\frac{1}{(1-\xi_{s}(\textbf{q}))}\frac{\hat{\chi}^{0}(\textbf{q},0)\hat{P}(\textbf{q})}{(1-\phi_{1}(\textbf{q}))(1-\phi_{2}(\textbf{q}))},
\end{eqnarray}
where $\xi_s(\textbf{q})$, $\phi_{1}(\textbf{q})$, and $\phi_{2}(\textbf{q})$ are the maximum and other eigenvalues of $\hat{U}\hat{\chi}^{0}(\textbf{q},0)$.
$\hat{P}(\textbf{q})$ is the adjugate matrix of $\hat{I}-\hat{U}\hat{\chi}^{0}(\textbf{q},0)$.
The eigenvalues of $\hat{U}\hat{\chi}^{0}(\textbf{q},0)$ is smaller than $1$ in the paramagnetic regime.
When $\xi_{s}(\textbf{q}) \rightarrow 1$, the spin susceptibility $\hat{\chi}^{s}(\textbf{q},0)$ diverges and induces a magnetic order, corresponding to the wavenumber $\textbf{q}$.

Within the framework of linear response theory, the Knight shift, $K$, and the spin-lattice relaxation rate, $(1/T_{1}T)$, for the orbital $\alpha$ are given by\cite{S.Katayama2009}
\begin{eqnarray}
\label{KS_formula}
K_{\alpha}&\propto& \sum_{\beta}{\rm Re}\left[\chi^{zz}_{\alpha\beta}(\textbf{0},0)\right],
\end{eqnarray}
\begin{eqnarray}
\label{T1T_formula}
\left(1/T_{1}T\right)_{\alpha}&\propto& \lim_{\omega \rightarrow +0}\left[\frac{1}{N_L}\sum_{\textbf{q}}\frac{{\rm Im}\chi^{\pm}_{\alpha\alpha}(\textbf{q},\omega)}{\omega}\right]
\end{eqnarray}
Here, note that Eq. \ref{chiS_RPA} is satisfied.
According to Eqs. \ref{xi_and_chiS} and \ref{T1T_formula}, all the $\textbf{q}$ components for which $\xi_{s}(\textbf{q})$ becomes close to unity make a dominant contribution to $1/T_{1}T$ because they lead to a large value in the spin susceptibility.
By contrast, the Knight shift is solely affected by the $\textbf{q}$=$\textbf{0}$ component that satisfies $\xi_{s}(\textbf{q})\sim 1$ (see Eq. \ref{KS_formula}).

Because the spin susceptibility in the real-frequency representation $\hat{\chi}^{s}(\textbf{q},\omega)$ is necessary for $1/T_{1}T$, we obtain $\hat{\chi}^{s}(\textbf{q},\omega)$ by performing an analytic continuation of Eq. \ref{chiS_RPA}. In this way, we use the real-frequency representation depending on the physical quantities.

\section{result in the absence of $U$}
In this section, we calculate the electronic state, spin susceptibility, Knight shift, and $1/T_{1}T$ in the absence of the repulsive interaction $U$.
Further, we find that the spin susceptibility in [Ni(dmdt)$_2$] greatly depends on the fragment orbitals and will explain the relationship between the spin susceptibilities and the wave functions.

\subsection{Electronic state and spin susceptibility in the absence of $U$}
\begin{figure}[htpb]
\begin{center}
\includegraphics[width=45mm]{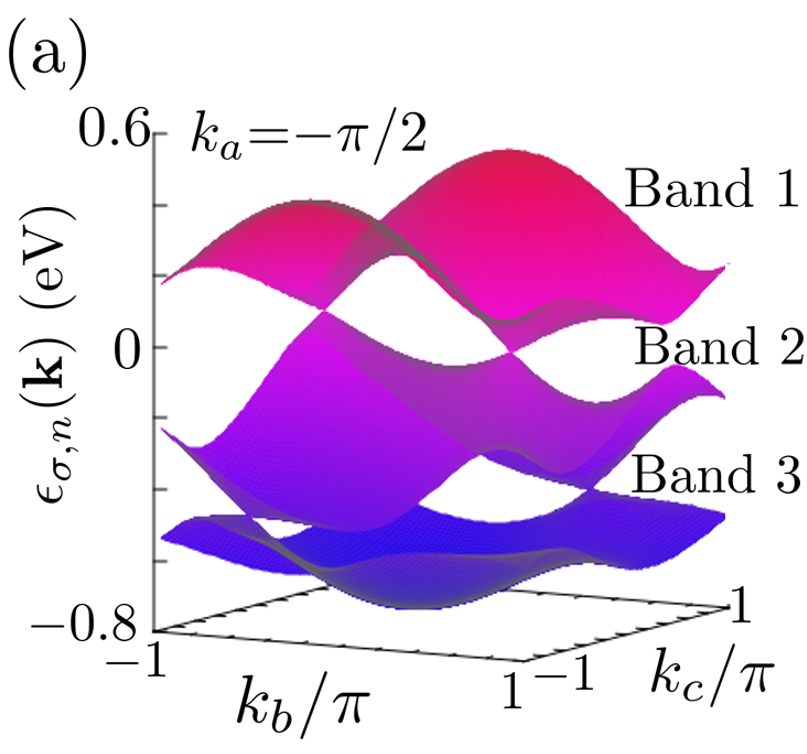}
\end{center}
\begin{center}
\includegraphics[width=65mm]{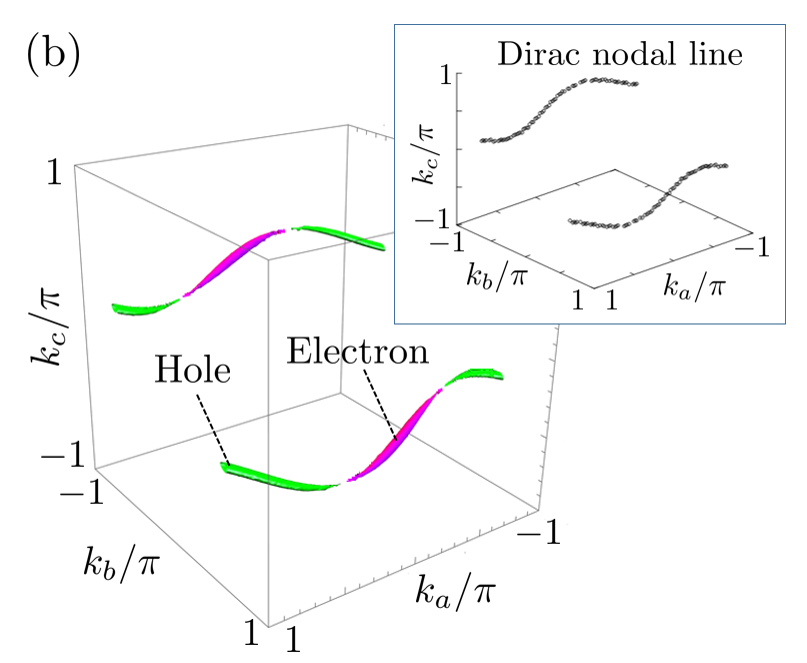}
\end{center}
\begin{center}
\includegraphics[width=85mm]{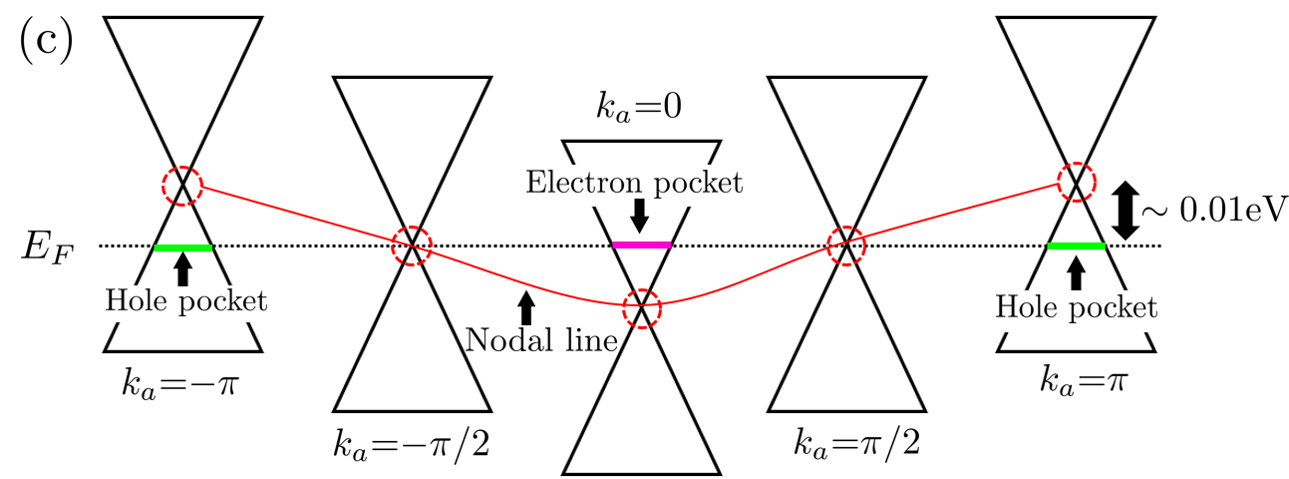}
\end{center}
\caption{ (a) Energy dispersion of [Ni(dmdt)$_2$] in the $k_b$--$k_c$ plane, where $k_a$=$-\pi/2$. Dirac cones exist between each pair of bands. Those between bands 1 and 2 draw the Dirac nodal line. (b) Fermi surface in the first Brillouin zone. The electron and hole pockets are drawn in magenta and green, respectively.
The inset shows the Dirac nodal line in the first Brillouin zone.
(c) Schematic of the relationship among the Dirac nodal line, Fermi surface, and wavenumber $k_a$. The red curved line shows the Dirac nodal line. The dotted transverse line shows $E_F$.
}
\label{band_FS}
\end{figure}

We first calculate the energy dispersion of [Ni(dmdt)$_2$] by diagonalizing the unperturbed Hamiltonian $\hat{H}^{0}_{\sigma}(\textbf{k})$ in Eq. \ref{eigen_Eq}.
The resulting energy bands are depicted in Fig. \ref{band_FS} (a), where the dispersion seen in the $k_b$--$k_c$ plane at $k_a$=$-\pi/2$ is shown.
Inside the 2D first Brillouin zone, two pairs of gapless Dirac cones appear between the first and second top bands near $E_F$ (the Band 1 and 2 in Fig. \ref{band_FS} (a), where $E_F$ is chosen as the energy origin) and between the second and third bands rather beneath $E_F$ (the Bands 2 and 3), where these band-crossing points are protected by space-inversion symmetry.

Dirac points between bands 1 and 2 in the $k_b$--$k_c$ plane draw the Dirac nodal lines in the $k_a$ direction. The inset of Fig. \ref{band_FS} (b) shows the Dirac nodal line in the first Brillouin zone.
Upon changing the momentum along the $k_a$ direction, the positions of the Dirac points near $E_F$ move in the $k_b$--$k_c$ plane and eventually form a pair of so-called Dirac nodal lines in the 3D Brillouin zone [see the inset of Fig. \ref{band_FS} (b)].
The band-crossing points accordingly move up and down across $E_F$ with changing the value of $k_a$ as illustrated in Fig. \ref{band_FS} (c), which generates electron and hole pockets around these nodal lines [see Fig. \ref{band_FS} (b), where the electron and hole pockets are illustrated as thin magenta and green strips, respectively].
Figure \ref{band_FS}(b) shows the Fermi surface in [Ni(dmdt)$_2$].
The energy scale of such Fermi pockets is approximately 0.010 eV.

\begin{figure}[htpb]
\begin{center}
\includegraphics[width=60mm]{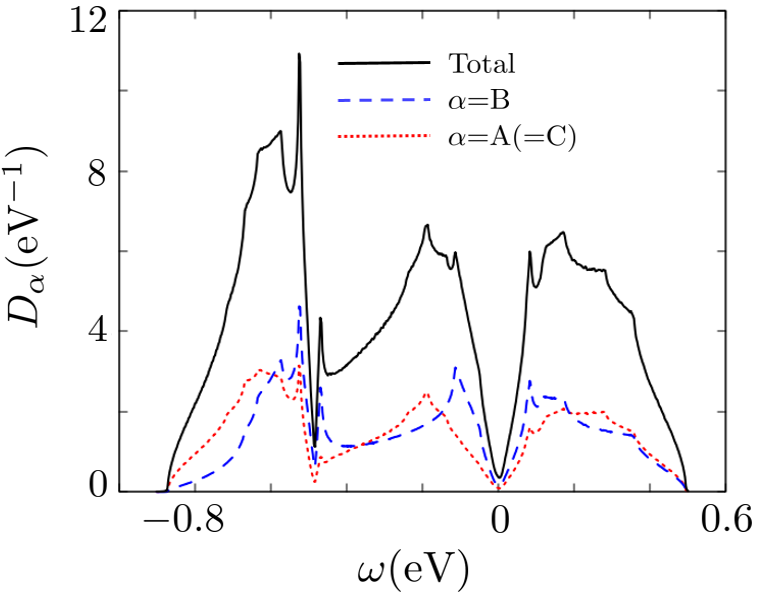}
\end{center}
\caption{
The local density of state $D_{\alpha}(\omega)$ for the fragment orbital $\alpha$=A, B, and C in the unit cell of [Ni(dmdt)$_2$]. The red dotted and blue dashed lines show $D_{A}(\omega)$(=$D_{C}(\omega)$) and $D_{B}(\omega)$, respectively. The black solid line shows the total density of states $D_{{\rm tot}}(\omega)$=$D_{A}(\omega)+D_{B}(\omega)+D_{C}(\omega)$. The integral value of $D_{{\rm tot}}(\omega)$ orver $\omega$ is equal to $6$ due to the three orbitals and the spins. Its unit is the number of electrons.
}
\label{DOS}
\end{figure}

The corresponding density of states (DOS), $D_{\rm tot}(\omega)$, was also
calculated, which is given by a sum of a fragment-orbital dependent DOS, $D_{\alpha}(\omega)$ [$\alpha$=A(=C) and B], that is given by
\begin{eqnarray}
\label{density of state}
&D_{\alpha}(\omega)=-\frac{1}{\pi N_{L}}\sum_{\textbf{k} , \sigma}{\rm Im}G^{R,0}_{\alpha\alpha , \sigma}(\textbf{k},\omega),
\end{eqnarray}
\begin{eqnarray}
\label{reterded G0}
&G^{R,0}_{\alpha\beta , \sigma}(\textbf{k},\omega)=\sum_{n}d_{\alpha,n,\sigma}(\textbf{k})d^{*}_{\beta,n,\sigma}(\textbf{k})\frac{1}{\omega-\epsilon_{\textbf{k},n,\sigma}+i\eta}.
\end{eqnarray}
Here, $\hat{G}^{R,0}(\textbf{k},\omega)$ is the non-interacting retarded Green's function, where $\eta$=$+0$.
The resulting $D_{\rm tot}(\omega)$ and $D_{\alpha}(\omega)$ are depicted in Fig. \ref{DOS}.
Note that the DOS has linear $\omega$ dependence near $E_{F}$ (corresponding to $\omega$=$0$ in Fig. \ref{DOS}) because the energy dispersion of this system near $E_F$ is close to that of a two-dimensional Dirac electron system, and the three-dimensional effect is only an addition of a small dispersion along a $k_{a}$ direction. The finite DOS at $E_F$ in this material is ascribed to the presence of the Fermi pockets induced by that $k_a$-axis dispersion.

\begin{figure}[htpb]
\begin{center}
\includegraphics[width=60mm]{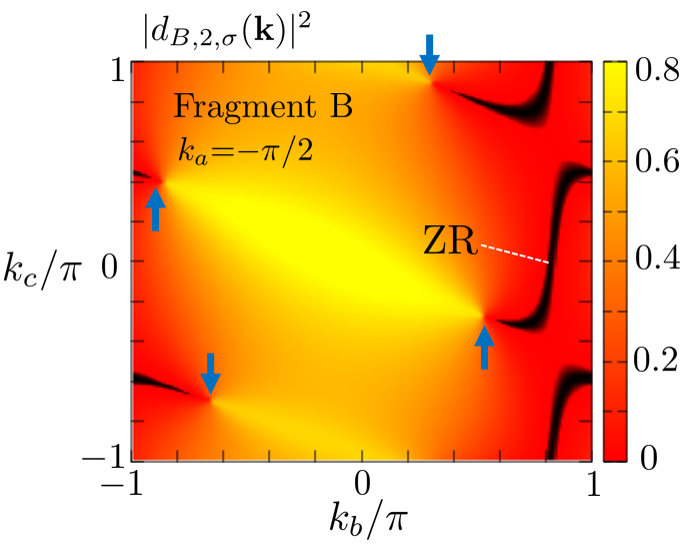}
\end{center}
\caption{ Absolute squared wave function of orbital B in band 2, $|d_{B,2,\sigma}(\textbf{k})|^{2}$, in the $k_b$--$k_c$ plane, where $k_a$=$-\pi/2$. The up arrows indicate the corresponding positions of the Dirac points formed between the Bands 1 and 2 in Fig. \ref{band_FS} (a), while the down arrows indicate those between the Bands 2 and 3. The color bar represents the magnitude of $|d_{B,2,\sigma}(\textbf{k})|^{2}$. }
\label{wavefunction}
\end{figure}

Before moving to the analysis of the spin susceptibility, further comments are needed on the fragment-orbital-dependent characters of this system.
Figure \ref{wavefunction} shows the momentum dependence of the squared wave function for the
orbital B projected onto the second top band [Band 2 in Fig. \ref{band_FS}(a)], $|d_{B,2,\sigma}(\textbf{k})|^{2}$, plotted as a function of $k_b$ and $k_c$ at $k_a$=$-\pi/2$.
Notably, the line segments that connect the positions of these Dirac points (illustrated
with black lines in Fig. \ref{wavefunction}) have a vanishing amplitude, $|d_{B,2,\sigma}(\textbf{k})|^{2}$=$0$, which we call the ``zero region'' in this paper.
By contrast, the wave functions of the orbitals A and C do not have such ZR. The presence of a similar ZR was previously found in other $n$-band ($n\geq 3$) Dirac electron systems, such as the organic conductors $\alpha$-(BEDT-TTF)$_2$I$_3$ ($n$=$4$)\cite{A.Kobayashi2013}.

Second, we calculate the non-interacting spin susceptibility $\hat{\chi}^{0}(\textbf{q},\omega)$ to elucidate spin fluctuations, which can be enhanced in this material.
\begin{figure}[htpb]
\begin{center}
\includegraphics[width=80mm]{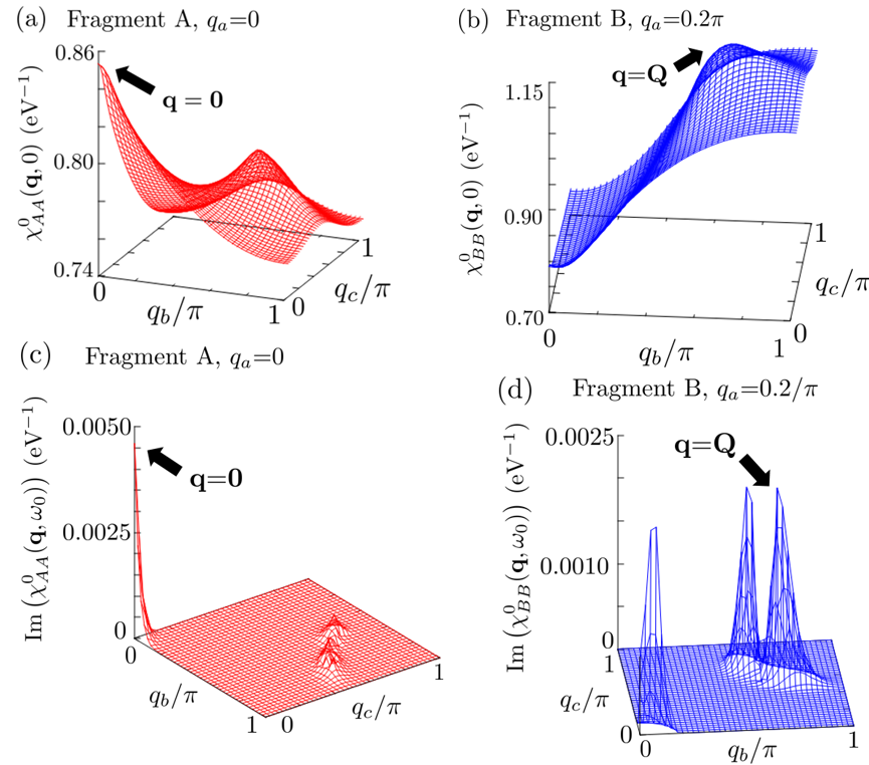}
\end{center}
\caption{ The momentum dependences of the diagonal elements of the spin susceptibility in the absence of $U$. (a) $\chi^{0}_{AA}(\textbf{q},0)$ in the $q_b$--$q_c$ plane, where $q_a$=$0$.
(b) $\chi^{0}_{BB}(\textbf{q},0)$ in the $q_b$--$q_c$ plane, where $q_a$=$0.2\pi$.
(c) ${\rm Im}[\chi^{0}_{AA}(\textbf{q},\omega_{0})]$ in the $q_b$--$q_c$ plane, where $q_a$=$0$.
(d) ${\rm Im}[\chi^{0}_{BB}(\textbf{q},\omega_{0})]$ in the $q_b$--$q_c$ plane, where $q_a$=$0.2\pi$.
The temperature $T$=$0.003$ eV.
}
\label{chi0}
\end{figure}
Figure \ref{chi0} (a), (b), (c), and (d) show the diagonal elements of the non-interacting spin susceptibility $\chi^{0}_{AA}(\textbf{q},0)$, $\chi^{0}_{BB}(\textbf{q},0)$, ${\rm Im}[\chi^{0}_{AA}(\textbf{q},\omega_{0})]$, and ${\rm Im}[\chi^{0}_{BB}(\textbf{q},\omega_{0})]$, respectively, at $T$=$0.003$ eV. These quantities slightly increase with raising temperature for $\textbf{q}$=$\textbf{0}$, while the magnitude relation $\chi^{0}_{AA}(\textbf{q}\sim \textbf{0},0)$$<$$\chi^{0}_{BB}(\textbf{q}\sim \textbf{Q},0)$ and ${\rm Im}\left[\chi^{0}_{AA}(\textbf{q}\sim \textbf{0},\omega_{0})\right]$$>$${\rm Im}\left[\chi^{0}_{BB}(\textbf{q}\sim \textbf{Q},\omega_{0})\right]$ do not change with temperature.
Here, $\chi^{0}_{\alpha\alpha}(\textbf{q},0)$ is a real number. We set $\omega_{0}$=$0.001$ eV because the imaginary part of the spin susceptibility at the infinitesimal frequency is essential to solve Eq. \ref{T1T_formula}. One of $q_a$, $q_b$, and $q_c$ must be fixed to show the spin susceptibilities in the three-dimensional figures. We fix $q_a$=$0$ in Fig. \ref{chi0} (a) and (c) and $q_a$=$0.2\pi$ in Fig. \ref{chi0} (b) and (d) because $\chi^{0}_{AA}(\textbf{q},0)$ and $\chi^{0}_{BB}(\textbf{q},0)$ have the maximum value at the commensurate wavenumber $\textbf{q}$=$\textbf{0}$ and the incommensurate wavenumber $\textbf{q}$=$\textbf{Q}$=$(0.20\pi,0.73\pi,0.58\pi)$ at $T$=$0.003$ eV, respectively.
We define $\textbf{Q}$ as the wavenumber at which $\chi^{0}_{BB}(\textbf{q},0)$ has the maximum value.
Further, $\textbf{Q}$ varies slightly with temperature.
$\chi^{0}_{CC}(\textbf{q},\omega)$ is equivalent to $\chi^{0}_{AA}(\textbf{q},\omega)$ owing to space-inversion symmetry.
In Fig. \ref{chi0} (a) and (c), $\chi^{0}_{AA}(\textbf{q},0)$ and ${\rm Im}[\chi^{0}_{AA}(\textbf{q},\omega_{0})]$ have their maximum value at $\textbf{q}$=$\textbf{0}$. In Fig. \ref{chi0} (b) and (d), $\chi^{0}_{BB}(\textbf{q},0)$ and ${\rm Im}[\chi^{0}_{BB}(\textbf{q},\omega_{0})]$ have their maximum value at $\textbf{q}$=$\textbf{Q}$. In addition, ${\rm Im}[\chi^{0}_{BB}(\textbf{q},\omega_{0})]$ has some peaks other than that at $\textbf{q}$=$\textbf{Q}$.

\begin{figure}[htpb]
\begin{center}
\includegraphics[width=85mm]{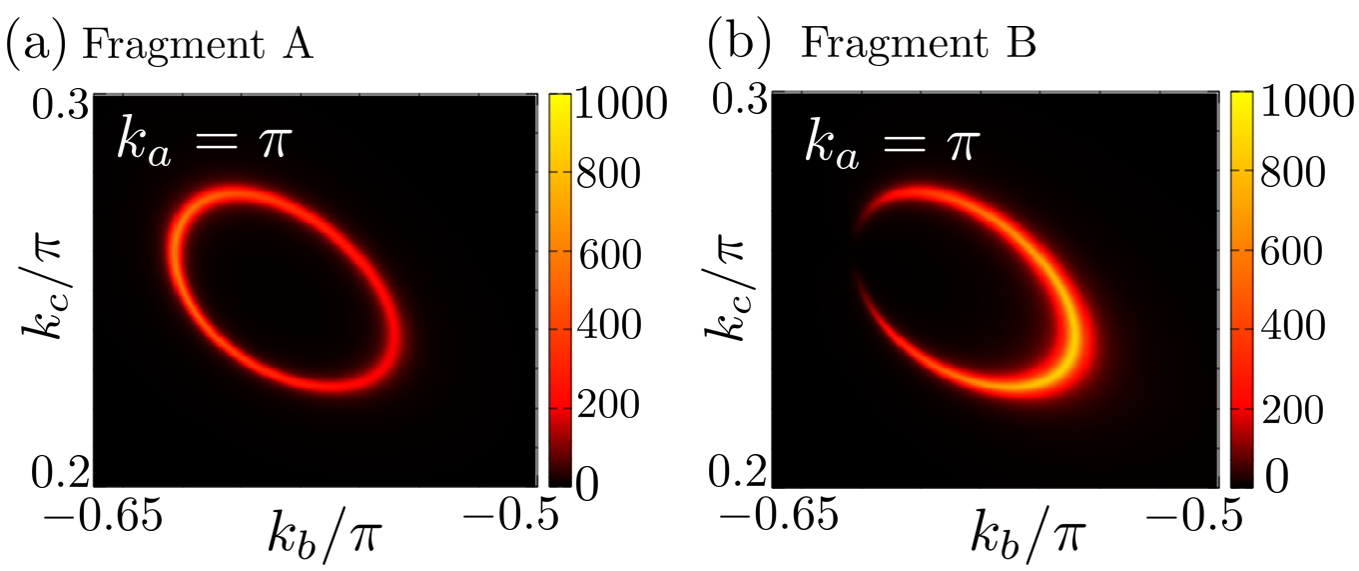}
\end{center}
\caption{
(a) Spectral weight $\rho_{A}(\textbf{k},0)$. It is not zero on the Fermi surface. (b) Spectral weight $\rho_{B}(\textbf{k},0)$. A part of it is zero because of a ZR. In the both figures, $k_{a}$=$\pi$, $-0.65\pi<k_{b}<-0.50\pi$, and $0.20\pi<k_{c}<0.30\pi$.
Color bars represent the magnitude of the spectral weights $\rho_{\alpha}(\pi,k_b,k_c,0)$ and $\rho_{\alpha}(0.8\pi,k_b,k_c,0)$. The yellow region indicates that the spectral weight at $E_{F}$ is high.
}
\label{spectral_weight}
\end{figure}

The difference between $\chi^{0}_{AA}(\textbf{q},0)$ and $\chi^{0}_{BB}(\textbf{q},0)$ implies that [Ni(dmdt)$_2$] has two candidates for magnetic order that can be induced in the bulk. They are the $\textbf{q}$=$\textbf{0}$ magnetic order and SDW.
To explain the mechanism of the fragment-orbital-dependent spin susceptibility, we calculate the spectral weights on the Fermi surface. The spectral weight is given by
\begin{equation}
\label{spectrum}
\rho_{\alpha}(\textbf{k},\omega)=-\frac{1}{\pi}{\rm Im}G^{R,0}_{\alpha\alpha}(\textbf{k},\omega).
\end{equation}
The spin index $\sigma$ is omitted in Eq. \ref{spectrum}, and $\textbf{k}$=$(k_a,k_b,k_c)$ is the wavenumber.
Eq. \ref{spectrum} with $\omega$=$0$ yields the spectral weight at $E_{F}$. The spectral weight shows the weights of the respective fragment orbitals for the energy $\omega$ and the wavenumber \textbf{k} because $G^{R,0}_{\alpha\alpha}(\textbf{k},\omega)$ in Eq. \ref{spectrum} contains the absolute square of the wave function $|d_{\alpha,n}(\textbf{k})|^{2}$.
Furthermore, the relationship $\frac{1}{N_L}\sum_{\textbf{k}}\rho_{\alpha}(\textbf{k},\omega)$=$D_{\alpha}(\omega)$ is satisfied.
Figure \ref{spectral_weight} (a) and (b) show the spectral weight on the cross-section where the Fermi surface in Fig. \ref{band_FS}(b) is cut on the $k_a$=$\pi$ plane, which corresponds to the hole pocket. We set $k_a$=$\pi$ because the hole pockets are important for $\chi^{0}_{BB}(\textbf{q},0)$, as we discuss below. Figure \ref{spectral_weight} (a) and (b), respectively, show $\rho_{A}(\textbf{k},0)$ and $\rho_{B}(\textbf{k},0)$ in the $k_{b}$--$k_{c}$ plane. In both figures, $-0.65\pi<k_{b}<-0.50\pi$ and $0.20\pi<k_{c}<0.30\pi$.
The spectral weight of A is not zero on the Fermi surface, but that of B has the appearance of a crescent moon because of a ZR. This difference in spectral weights results in the fragment-orbital-dependent spin susceptibilities.

\begin{figure}[htpb]
\begin{center}
\includegraphics[width=90mm]{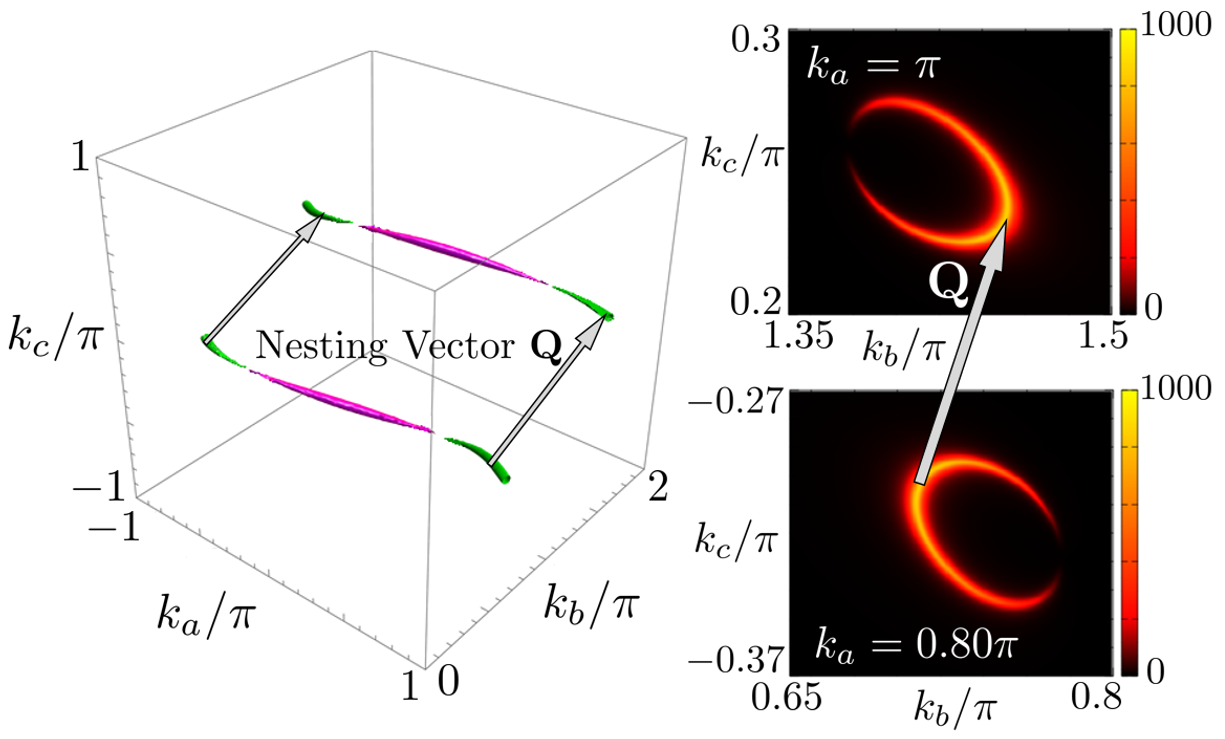}
\end{center}
\caption{ Nesting vector $\textbf{Q}$ in the momentum space.
$\textbf{Q}$ connects the regions where the spectral weight of orbital B is high (in the right figure).
Color bars represent the magnitude of the spectral weight $\rho_{B}(\pi,k_b,k_c,0)$. The electron and hole pockets are drawn in magenta and green, respectively (in the left figure).
}
\label{nesting_Q2}
\end{figure}

After performing summation over $i\tilde{\omega}_{m}$ in Eq. \ref{chi0_formula}, the non-interacting spin susceptibility is written as
\begin{equation}
\label{f-f/e-e}
\begin{split}
&\chi^{0}_{\alpha,\beta}(\textbf{q},i\omega_{l})=-\frac{1}{N_L}\sum_{\textbf{k},m,n}\frac{f_{\textbf{k}+\textbf{q},m}-f_{\textbf{k},n}}{\epsilon_{\textbf{k}+\textbf{q},m}-\epsilon_{\textbf{k},n}-i\omega_{l}}\\
&\times d_{\alpha,m}(\textbf{k}+\textbf{q})d^{*}_{\beta,m}(\textbf{k}+\textbf{q})
d_{\beta,n}(\textbf{k})d^{*}_{\alpha,n}(\textbf{k}),
\end{split}
\end{equation}
where the spin index $\sigma$ is omitted.
The terms in which the denominator is close to $0$ and the numerator is close to $1$ in Eq. \ref{f-f/e-e} contribute to $\chi^{0}_{\alpha \beta}(\textbf{q},0)$. Such terms are given by the wavenumber $\textbf{k}+\textbf{q}$ and $\textbf{k}$ near the Fermi surface. Thus, the vector $\textbf{q}$ connecting the Fermi surface is important for the spin susceptibility.
This wavenumber is called the nesting vector.
$\textbf{q}$=$\textbf{Q}$ in Fig. \ref{chi0} corresponds to the nesting vector $\textbf{Q}$ in Fig. \ref{nesting_Q2}. $\textbf{q}$=$\textbf{Q}$ connects the regions where the spectral weight of orbital B is high on the hole pockets. \textbf{Q} is the wavenumber at which $\chi^{0}_{BB}(\textbf{q},0)$ has the maximum value. The nesting between the electron pockets is relatively weaker than that between the hole pockets.
Note that we solved Eq. \ref{chi0_formula} using a fast Fourier transform, instead of solving Eq. \ref{f-f/e-e}.

\begin{figure}[htpb]
\begin{center}
\includegraphics[width=70mm]{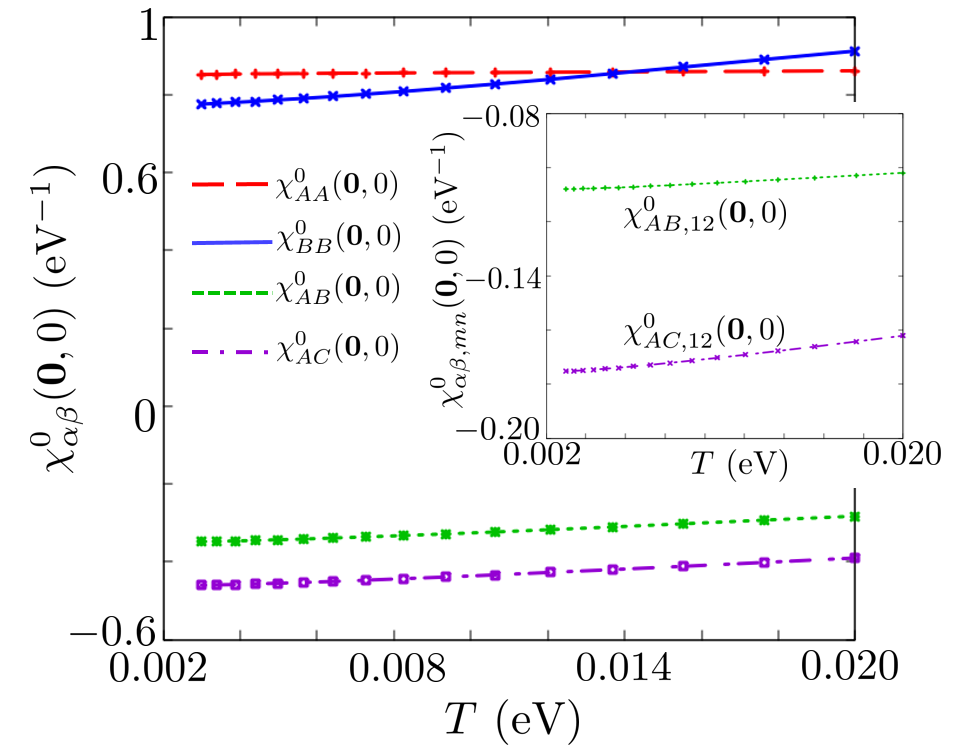}
\end{center}
\caption{ Temperature dependence of $\chi^{0}_{\alpha\beta}(\textbf{0},0)$. The red dashed, blue solid, green dotted, and purple chain lines represent $\chi^{0}_{AA}(\textbf{0},0)$, $\chi^{0}_{BB}(\textbf{0},0)$, $\chi^{0}_{AB}(\textbf{0},0)$, and $\chi^{0}_{AC}(\textbf{0},0)$, respectively. The inset shows the band-dependent spin susceptibilities $\chi^{0}_{AB,12}(\textbf{0},0)$ and $\chi^{0}_{AC,12}(\textbf{0},0)$ by the green dotted and purple chain lines, respectively.}
\label{chi0_00}
\end{figure}

Next, we calculate the temperature dependence of the matrix elements of $\hat{\chi}^{0}(\textbf{0},0)$ using Eq. \ref{chi0_formula} because it is essential to the following calculations. They are real numbers because $\textbf{q}$=$0$ and $\omega$=$0$. $\chi^{0}_{AA}(\textbf{0},0)$=$\chi^{0}_{CC}(\textbf{0},0)$, $\chi^{0}_{AB}(\textbf{0},0)$=$\chi^{0}_{BA}(\textbf{0},0)$=$\chi^{0}_{BC}(\textbf{0},0)$=$\chi^{0}_{CB}(\textbf{0},0)$, and $\chi^{0}_{AC}(\textbf{0},0)$=$\chi^{0}_{CA}(\textbf{0},0)$ are satisfied owing to space-inversion symmetry and time-reversal symmetry.
Figure \ref{chi0_00} shows the temperature dependence of the matrix elements of $\hat{\chi}^{0}(\textbf{0},0)$.
The diagonal element $\chi^{0}_{AA}(\textbf{0},0)$ is almost constant. Furthermore, $\chi^{0}_{BB}(\textbf{0},0)$ decreases slowly with decreasing temperature.
However, the off-diagonal elements $\chi^{0}_{AB}(\textbf{0},0)$ and $\chi^{0}_{AC}(\textbf{0},0)$ are negative and decrease as the temperature decreases.
To show what determines the sign of $\chi^{0}_{AB}(\textbf{0},0)$ and $\chi^{0}_{AC}(\textbf{0},0)$, we define the band-dependent spin susceptibility as
\begin{equation}
\label{chi0_band}
\begin{split}
&\chi^{0}_{\alpha\beta,mn}(\textbf{q},i\omega_{l}) \\
&=-\frac{T}{N_L}\sum_{\textbf{k},l'}G^{0}_{\alpha\beta,m}(\textbf{k}+\textbf{q},i\tilde{\omega}_{l'}+i\omega_{l})G^{0}_{\beta\alpha,n}(\textbf{k},i\tilde{\omega}_{l'}),
\end{split}
\end{equation}
\begin{equation}
\label{Green_band}
\begin{split}
G^{0}_{\alpha\beta,m,\sigma}(\textbf{k},i\tilde{\omega}_{l})=d_{\alpha,m,\sigma}(\textbf{k})d^{*}_{\beta,m,\sigma}(\textbf{k})\frac{1}{i\tilde{\omega}_{l}-\epsilon_{\textbf{k},m,\sigma}},
\end{split}
\end{equation}
where $m$ and $n$ are the band indices.
Eq. \ref{chi0_band} satisfies $\sum_{m,n}\chi^{0}_{\alpha\beta,mn}(\textbf{q},i\omega_{l})$= $\chi^{0}_{\alpha\beta}(\textbf{q},i\omega_{l})$, where $\chi^{0}_{\alpha\beta}(\textbf{q},i\omega_{l})$ is given by Eq. \ref{chi0_formula}.
The inset of Fig. \ref{chi0_00} shows the temperature dependence of $\chi^{0}_{AB,12}(\textbf{0},0)$ and $\chi^{0}_{AC,12}(\textbf{0},0)$. They are negative. Such terms render the off-diagonal elements of spin susceptibility, $\chi^{0}_{AB}(\textbf{0},0)$ and $\chi^{0}_{AC}(\textbf{0},0)$, negative.

\subsection{Knight shift and $1/T_{1}T$ in the absence of $U$}
\begin{figure}[htpb]
\begin{center}
\includegraphics[width=60mm]{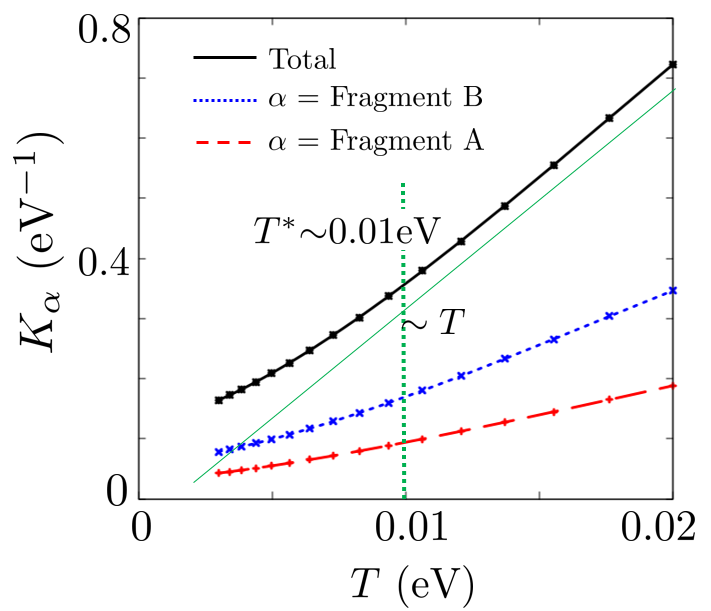}
\end{center}
\caption{ Temperature dependence of $K_{\alpha}$ in the absence of $U$.
The red dashed and blue dotted lines show $K_{A}$ and $K_{B}$, respectively. The black solid line shows the total Knight shift $K_{total}$=$K_{A}+K_{B}+K_{C}$.
The dotted longitudinal line indicates $T$=$T^{*}$$\sim$$0.01$ eV. The green thin line which is proportional to $T$ is drawn for the eye guide.
}
\label{KS0}
\end{figure}

We calculate the Knight shift in the absence of $U$ using Eq. \ref{KS_formula}. The fragment orbital components of the Knight shift can be given by $K_{A}$=$\chi^{0}_{AA}(\textbf{0},0)+\chi^{0}_{AB}(\textbf{0},0)+\chi^{0}_{AC}(\textbf{0},0)$ and $K_B$=$\chi^{0}_{BB}(\textbf{0},0)+2\chi^{0}_{AB}(\textbf{0},0)$ using the spin susceptibilities in Fig. \ref{chi0_00} due to the space-inversion symmetry.
Figure \ref {KS0} shows the Knight shift for the fragment orbitals A(=C) and B in the absence of $U$.
$T$=$T^{*}$$\sim$$0.01$eV is the energy scale where the Fermi surface affects the Knight shift and $1/T_{1}T$. $T^{*}$$\sim$$0.01$eV is consistent with the energy scale of the Fermi surface. For temperatures higher than $T^{*}$, the Knight shift and $1/T_{1}T$ are affected by the linear energy dispersion.
The Knight shift of the Dirac electron system in the absence of the interaction is given by
$K_{\alpha} \simeq \int^{\infty}_{-\infty} D_{\alpha}(\omega) \left(-\frac{df(\omega)}{d\omega}\right)d\omega$.\cite{S.Katayama2009}
$f(\omega)$ is the Fermi distribution function for the energy $\omega$.
Because $D_{\alpha}(\omega) \propto \omega$ near the Femi energy in Fig. \ref{DOS},
$K_{\alpha}$ is proportional to $T$ for $T\gtrsim T^{*}$ in Fig. \ref{KS0}.
In the two-dimensional Dirac electron system, the Knight shift is proportional to $T$ at low temperature and becomes zero at $T$=$0$ because DOS is zero at $E_{F}$.
However, $K_{\alpha}$ in this material is not proportional to $T$ for $T\lesssim T^{*}$ because $D_{\alpha}(0) \neq 0$ in Fig. \ref{DOS}. This is the effect of the Fermi surface. The magnitude relationship $K_{B}> K_{A}$ results from $D_{B}(\omega) > D_{A}(\omega)$ near the Fermi energy.

\begin{figure}[htpb]
\begin{center}
\includegraphics[width=60mm]{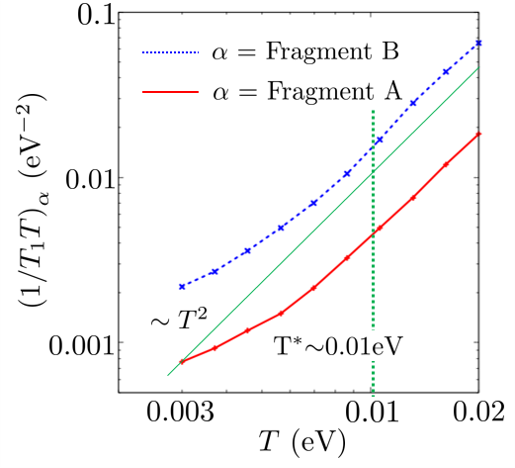}
\end{center}
\caption{
Temperature dependence of $(1/T_{1}T)_{\alpha}$ in the absence of $U$. The red solid and blue dotted lines show $(1/T_{1}T)_{A}$ and $(1/T_{1}T)_{B}$, respectively. The dashed longitudinal line indicates $T$=$T^{*}$$\sim$$0.01$ eV. The green thin line which is proportional to $T^{2}$ is drawn for the eye guide.
}
\label{T1T0}
\end{figure}

Next, we calculate the spin-lattice relaxation rate $1/T_{1}T$ in the absence of $U$ using Eq. \ref{T1T_formula}.
$1/T_{1}T$ is determined by $\sum_{\textbf{q}}{\rm Im}[\chi^{0}_{\alpha\alpha}(\textbf{q},\omega_{0})]$.
${\rm Im}[\chi^{0}_{AA}(\textbf{q},\omega_{0})]$ and ${\rm Im}[\chi^{0}_{BB}(\textbf{q},\omega_{0})]$ are shown in Fig. \ref{chi0}.
Fig. \ref{T1T0} shows the temperature dependence of $1/T_{1}T$ in the absence of $U$.
$(1/T_{1}T)_{\alpha}$ is proportional to $T^{2}$ for $T\gtrsim T^{*}$ in Fig. \ref{KS0} because the $1/T_{1}T$ of the Dirac electron system is given by $(1/T_{1}T)_{\alpha} \simeq \int^{\infty}_{-\infty} [D_{\alpha}(\omega)]^{2} \left(-\frac{df(\omega)}{d\omega}\right)d\omega$.\cite{S.Katayama2009} It means that $\textbf{q}$=$\textbf{0}$ components of the imaginary parts of the spin susceptibilities contribute to $1/T_1 T$.
For $T\lesssim T^{*}$, $(1/T_{1}T)_{\alpha}$ is not proportional to $T^{2}$, because of the Fermi surface.

\section{result in the presence of $U$}
In this section, we calculate spin fluctuations, the Knight shift, and $1/T_{1}T$ in the presence of $U$. It is shown that electron correlation effects are important for explaining the observed
temperature dependence of the Knight shift and $1/T_{1}T$ reported by $^{13}$C-NMR experiments.\cite{T.Sekine.Private}
According to Eq. \ref{xi_and_chiS}, the Stoner factor $\xi_{s}(\textbf{q})$ that has a value close to unity gives major contributions to the spin susceptibility and therefore to the Knight shift and $1/T_{1}T$.
$\xi_{s}(\textbf{q}=\textbf{0})\approx 1$ contributes to the Knight
shift (Eq. \ref{KS_formula}) and $1/T_{1}T$. $\xi_{s}(\textbf{q}=\textbf{Q})\approx 1$ does not contribute to the Knight shift but to $1/T_{1}T$  (Eq. \ref{T1T_formula}).

\subsection{Spin fluctuations}
\begin{figure}[htpb]
\begin{center}
\includegraphics[width=60mm]{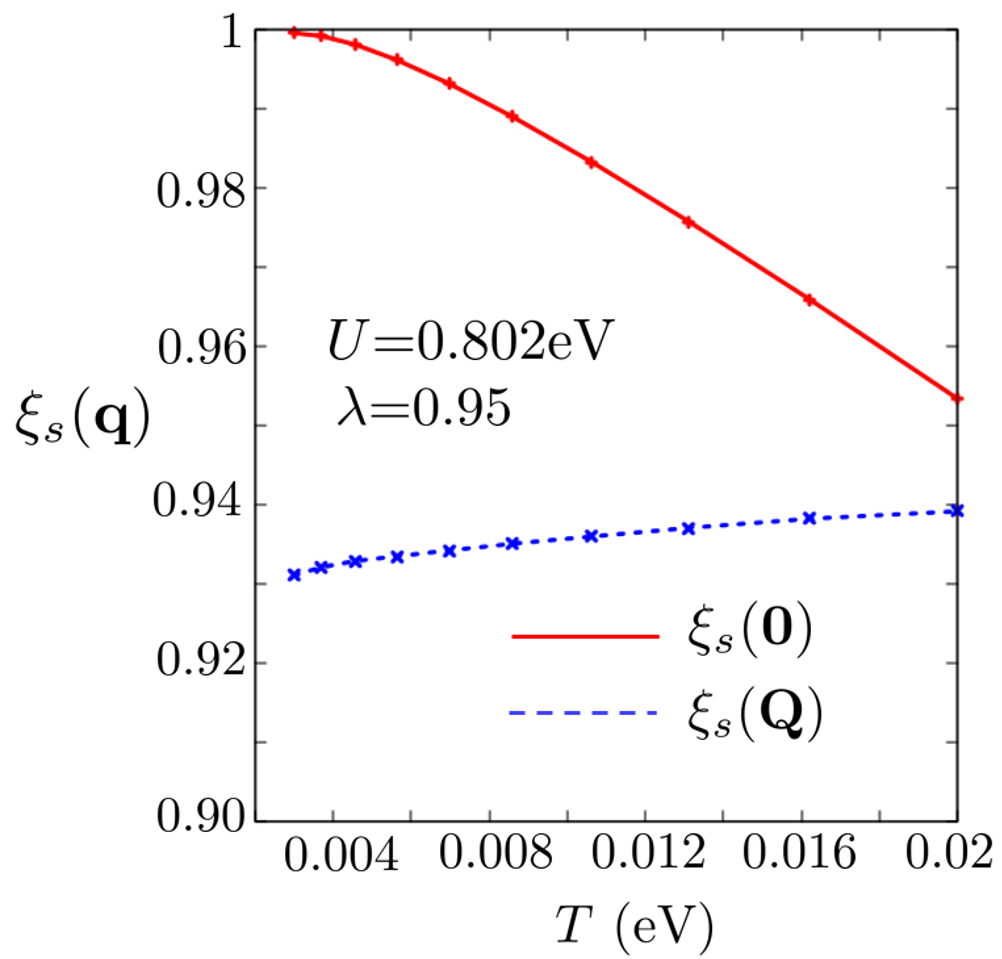}
\end{center}
\caption{ Temperature dependence of the Stoner factors $\xi_s(\textbf{0})$ and $\xi_s(\textbf{Q})$. The red solid and blue dotted lines show $\xi_{s}(\textbf{0})$ and $\xi_{s}(\textbf{Q})$, respectively.}
\label{Stoner}
\end{figure}

We calculate the temperature dependence of the Stoner factor $\xi_{s}(\textbf{q})$ because $\xi_{s}(\textbf{q})$ is an important measurement of spin fluctuations as discussed in Section I\hspace{-.1em}I.
Figure \ref{Stoner} shows the temperature dependence of $\xi_{s}(\textbf{q})$, where $U$=$0.802$ and $\lambda$=$0.95$. $\lambda$=$0.95$ was obtained in the calculation of the screened Coulomb interaction using \textsc{respack}.
In the case of $\lambda$=$0.95$, $\xi_{s}(\textbf{0})$ is maximum in the momentum space and increases as $T$ decreases. The combination of $\lambda$=$0.95$ and $U$=$0.802$ yields $\xi_{s}(\textbf{0})$=$0.999$ at $T$=$0.003$ eV. On the other hand, $\xi_{s}(\textbf{Q})$ decreases slowly with decreasing $T$ and is less than $\xi_{s}(\textbf{0})$.
$\textbf{Q}$ is the wavenumber at which $\chi^{0}_{BB}(\textbf{q},0)$ has the maximum value.
$\chi^{s}_{BB}(\textbf{q},0)$ also has the maximum value at $\textbf{q}$=$\textbf{Q}$.
The magnitude relationship $\xi_{s}(\textbf{0})>\xi_{s}(\textbf{Q})$ implies that the $\textbf{q}$=$\textbf{0}$ magnetic order is easier to induce than SDW at $\textbf{q}$=$\textbf{Q}$.
At a low temperature of $T$$\lesssim$$0.003$eV, it is difficult for $\xi_{s}(\textbf{q})$ to converge in the numerical calculation because enormous Matsubara frequencies and wavenumbers are required.

\begin{figure}[htpb]
\begin{center}
\includegraphics[width=60mm]{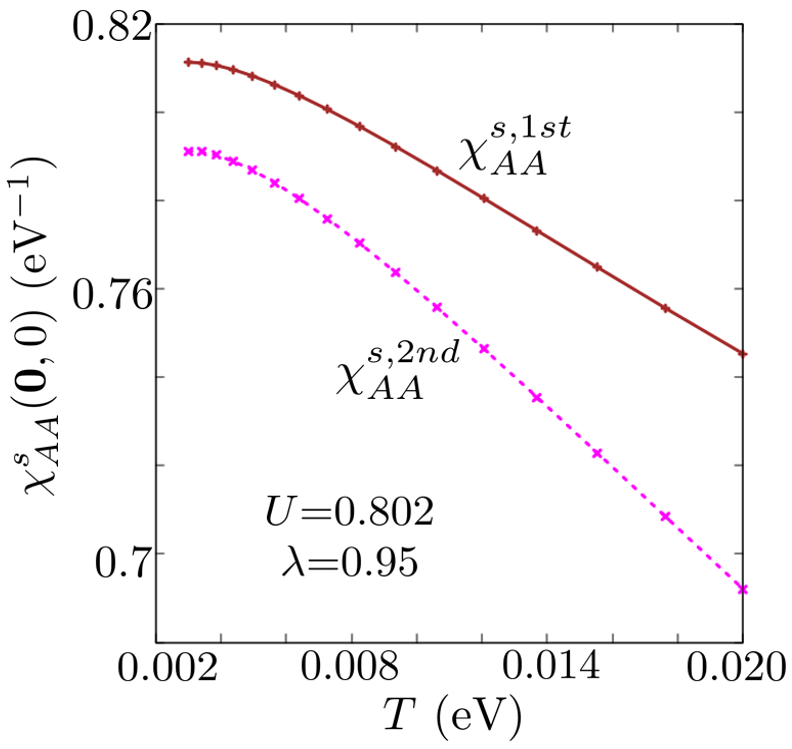}
\end{center}
\caption{ First- and second-order perturbation terms of $\chi^{s}_{AA}(\textbf{0},0)$ shown by the brown solid and magenta dotted line, respectively. The horizontal axis represents the temperature.}
\label{1+2}
\end{figure}

We explain why $\xi_{s}(\textbf{0})$ increases at low temperature.
In fact, $\xi_{s}(\textbf{q})$ isn't directly determined by the DOS because the maximum eigenvalue of $\hat{U}\hat{\chi}^{0}(\textbf{q},0)$ contains the products of the $\chi^{0}_{\alpha\beta}(\textbf{q},0)$. It means that the electron correlation effect is important.
We calculate the first- and second-order perturbation terms in Eq. (A) and (B) in Fig. \ref{diagram}. In this study, Eq. (A) and (B) in Fig. \ref{diagram} are equivalent.
Their matrix elements $\chi^{s,1st}_{\alpha\beta}(\textbf{q},i\omega_{l})$ and $\chi^{s,2nd}_{\alpha\beta}(\textbf{q},i\omega_{l})$ are written as
\begin{eqnarray}
\label{chiS_1st}
\chi^{s,1st}_{\alpha\beta}(\textbf{q},i\omega_{l})&=&\sum_{\gamma}\chi^{0}_{\alpha\gamma}(\textbf{q},i\omega_{l})U_{\gamma\gamma}\chi^{0}_{\gamma\beta}(\textbf{q},i\omega_{l}),
\end{eqnarray}
\begin{eqnarray}
\label{chiS_2nd}
\chi^{s,2nd}_{\alpha\beta}(\textbf{q},i\omega_{l})&=&\sum_{\gamma,\gamma'}\chi^{0}_{\alpha\gamma}(\textbf{q},i\omega_{l})U_{\gamma\gamma}\chi^{0}_{\gamma\gamma'}(\textbf{q},i\omega_{l}) \\ \nonumber
&\times& U_{\gamma'\gamma'}\chi^{0}_{\gamma'\beta}(\textbf{q},i\omega_{l}).
\end{eqnarray}
They are the first- and second-order perturbation terms in RPA and correspond to the second and third terms in Eq. (A) or (B) in Fig. \ref{diagram}, respectively.
Figure \ref{1+2} shows the temperature dependence of $\chi^{s,1st}_{AA}(\textbf{0},0)$ and $\chi^{s,2nd}_{AA}(\textbf{0},0)$. They increase as $T$ decreases. Note that the zero-order perturbation term is identical to the red dashed line in Fig. \ref{chi0_00}.
Since the off-diagonal elements of $\hat{\chi}^{0}(\textbf{0},0)$ are negative and decrease as $T$ decreases in Fig. \ref{chi0_00}, their absolute squares increase as $T$ decreases. Thus, terms such as $\chi^{0}_{AB}(\textbf{0},0)U_{BB}\chi^{0}_{BA}(\textbf{0},0)$ in Eq. \ref{chiS_1st} and $\chi^{0}_{AC}(\textbf{0},0)U_{CC}\chi^{0}_{CC}(\textbf{0},0)U_{CC}\chi^{0}_{CA}(\textbf{0},0)$ in Eq. \ref{chiS_2nd} increase $\chi^{s,1st}_{AA}(\textbf{0},0)$ and $\chi^{s,2nd}_{AA}(\textbf{0},0)$ at low temperature.
The other higher-order perturbation terms behave similarly. Therefore, $\xi_{s}(\textbf{0})$ increases as $T$ decreases.

\begin{figure}[htpb]
\begin{center}
\includegraphics[width=65mm]{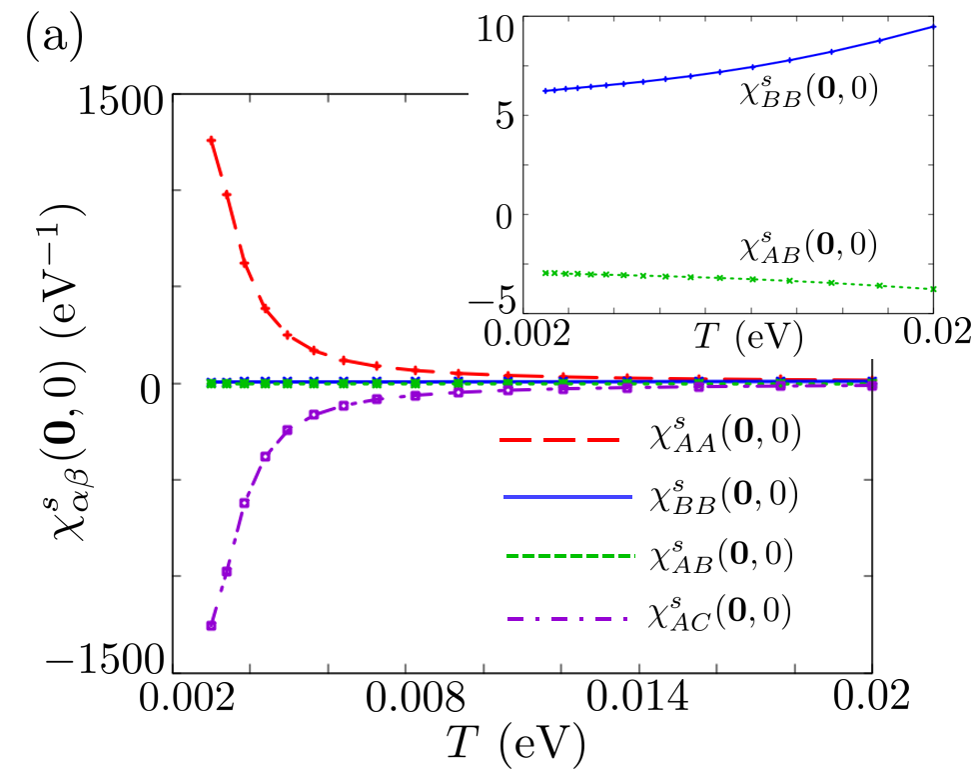}
\includegraphics[width=55mm]{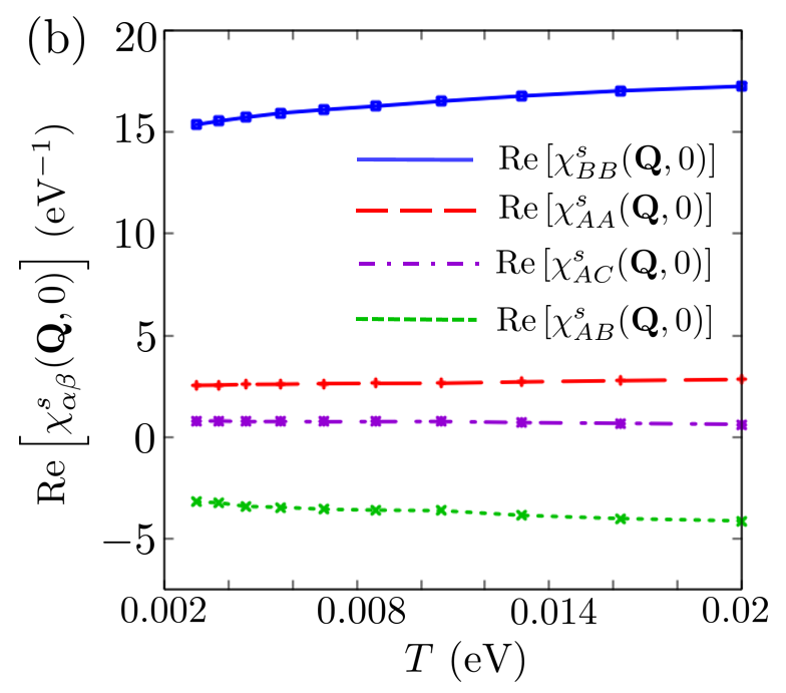}
\end{center}
\caption{ (a) Temperature dependence of $\chi^{s}_{AA}(\textbf{0},0)$, $\chi^{s}_{BB}(\textbf{0},0)$, $\chi^{s}_{AB}(\textbf{0},0)$, and $\chi^{s}_{AC}(\textbf{0},0)$ shown by the red dashed, blue solid, green dotted, and purple chain lines, respectively. The inset shows an enlarged view of the region around $\chi^{s}_{BB}(\textbf{0},0)$ and $\chi^{s}_{AB}(\textbf{0},0)$.
(b) Temperature dependence of ${\rm Re}\left[ \chi^{s}_{AA}(\textbf{Q},0) \right]$, ${\rm Re}\left[ \chi^{s}_{BB}(\textbf{Q},0) \right]$, ${\rm Re}\left[ \chi^{s}_{AB}(\textbf{Q},0) \right]$, and ${\rm Re}\left[ \chi^{s}_{AC}(\textbf{Q},0) \right]$. The combination of matrix elements and lines is the same as in (a).}
\label{chiS_00}
\end{figure}

Fragment orbitals A and C are sensitive to $\xi_{s}(\textbf{0})$. On the other hand, fragment orbital B is insensitive to $\xi_{s}(\textbf{0})$ but sensitive to $\xi_{s}(\textbf{Q})$.
We calculate the matrix elements of $\hat{\chi}^{s}(\textbf{0},0)$ and $\hat{\chi}^{s}(\textbf{Q},0)$ to exhibit this phenomenon using Eqs. \ref{chi0_formula} and \ref{chiS_RPA}.
Figure \ref{chiS_00}(a) shows the temperature dependence of $\chi^{s}_{AA}(\textbf{0},0)$, $\chi^{s}_{BB}(\textbf{0},0)$, $\chi^{s}_{AB}(\textbf{0},0)$, and $\chi^{s}_{AC}(\textbf{0},0)$. They are real numbers because $\textbf{q}$=$0$ and $\omega$=$0$. $\chi^{s}_{AA}(\textbf{0},0)$=$\chi^{s}_{CC}(\textbf{0},0)$, $\chi^{s}_{AB}(\textbf{0},0)$=$\chi^{s}_{BA}(\textbf{0},0)$=$\chi^{s}_{BC}(\textbf{0},0)$=$\chi^{s}_{CB}(\textbf{0},0)$, and $\chi^{s}_{AC}(\textbf{0},0)$=$\chi^{s}_{CA}(\textbf{0},0)$ are satisfied because of space-inversion symmetry and time-reversal symmetry. The inset shows an enlarged view of the region around $\chi^{s}_{BB}(\textbf{0},0)$ and $\chi^{s}_{AB}(\textbf{0},0)$. $\chi^{s}_{BB}(\textbf{0},0)$ and $\chi^{s}_{AB}(\textbf{0},0)$ are difficult to increase at low temperature, whereas $\chi^{s}_{AA}(\textbf{0},0)$ sharply increases. Moreover, $\chi^{s}_{AC}(\textbf{0},0)$ is negative and sharply decreases as $T$ decreases.
$\xi_{s}(\textbf{0})$ easily affects fragment orbitals A and C but not fragment orbital B.
The situation of $\chi^{s}_{AC}(\textbf{0},0)$$<$$0$ in Fig. \ref{chiS_00}(a) implies intra-molecular antiferromagnetic fluctuations. The negative off-diagonal elements of the spin susceptibility are due to the characteristic wave function of the Dirac nodal line system.

Figure \ref{chiS_00}(b) shows the temperature dependence of ${\rm Re}\left[ \chi^{s}_{AA}(\textbf{Q},0) \right]$, ${\rm Re}\left[ \chi^{s}_{BB}(\textbf{Q},0) \right]$, ${\rm Re}\left[ \chi^{s}_{AB}(\textbf{Q},0) \right]$, and ${\rm Re}\left[ \chi^{s}_{AC}(\textbf{Q},0) \right]$. ${\rm Re}\left[ \hat{\chi}^{s}(\textbf{Q},0) \right]$ reflects $\xi_{s}(\textbf{Q})$ and slowly varies with temperature. At $\textbf{q}$=$\textbf{Q}$, ${\rm Re}\left[ \chi^{s}_{BB}(\textbf{Q},0) \right]$ is the largest of all matrix elements and increases with temperature. Fragment orbital B is sensitive to $\xi_{s}(\textbf{Q})$. Because ${\rm Re}\left[ \chi^{s}_{AB}(\textbf{Q},0) \right]$$<$$0$ and ${\rm Re}\left[ \chi^{s}_{AC}(\textbf{Q},0) \right]$$>$$0$, the spins of the fragment orbitals B and A(C) within a molecule are inversely correlated.

\begin{figure}[htpb]
\begin{center}
\includegraphics[width=70mm]{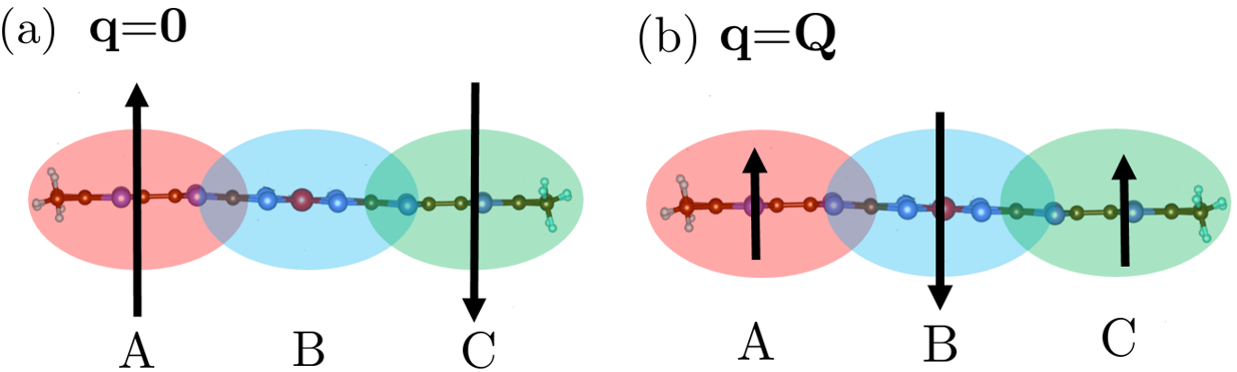}
\end{center}
\caption{
Schematic illustrations of spin polarization. (a) intra-molecular antiferromagnetic spin fluctuations linked to the $\textbf{q}$=$\textbf{0}$ response shown in Fig. \ref{chiS_00}(a). (b) Spin
correlations within a molecule given by the $\textbf{q}$=$\textbf{Q}$ response in Fig. 16(b).
}
\label{order-q}
\end{figure}

Figure \ref{order-q} schematically illustrates the spin polarization pictures that we
obtained from the calculated results in Fig. \ref{chiS_00} for the paramagnetic regime. Figure \ref{order-q}(a) corresponds to the case in Fig. \ref{chiS_00} (a), where we found intra-molecular antiferromagnetic spin fluctuations, which are commensurate($\textbf{q}$=$\textbf{0}$) between molecules. The solid arrows in Fig. \ref{order-q} (a) represents a tendency where an infinitesimally small downward local magnetic field at the orbital C(A) induces an upward spin polarization at the orbital A(C) by the linear response relation $M_{A(C)}$=$\chi^{s}_{AC(CA)}H_{C(A)}$, respectively [$M$ is the magnetization and $H$ is the infinitesimal magnetic field].
Figure \ref{order-q}(b), which is derived from Fig. \ref{chiS_00}(b), stand for the spin fluctuations within a molecule that are incommensurate (\textbf{q}=\textbf{Q}) between molecules. In this case, the spins at the orbitals A(=C) and B tend to be inversely correlated within a molecule.


\begin{figure}[htpb]
\begin{center}
\includegraphics[width=80mm]{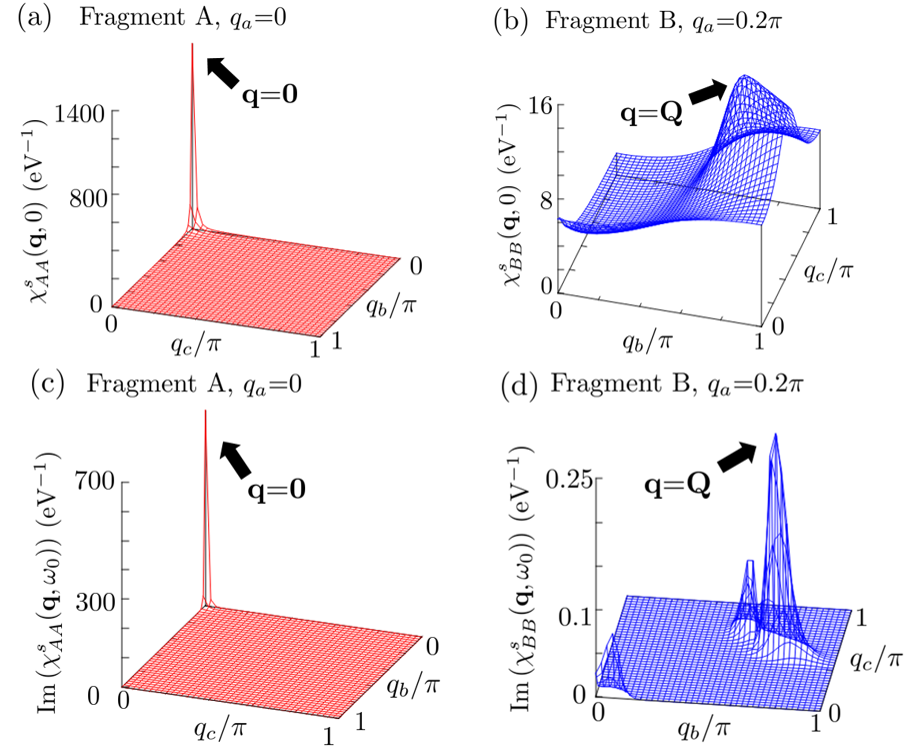}
\end{center}
\caption{ The momentum dependences of the diagonal elements of the spin susceptibility in the presence of $U$ (a) $\chi^{s}_{AA}(\textbf{q},0)$ in the $q_b$--$q_c$ plane, where $q_a$=$0$.
(b) $\chi^{s}_{BB}(\textbf{q},0)$ in the $q_b$--$q_c$ plane, where $q_a$=$0.2\pi$.
(c) ${\rm Im}[\chi^{s}_{AA}(\textbf{q},\omega_{0})]$ in the $q_b$--$q_c$ plane, where $q_a$=$0$.
(d) ${\rm Im}[\chi^{s}_{BB}(\textbf{q},\omega_{0})]$ in the $q_b$--$q_c$ plane, where $q_a$=$0.2\pi$.
The temperature $T$=$0.003$ eV.
}
\label{chiS}
\end{figure}

Next, we show the diagonal elements of the spin susceptibilities $\chi^{s}_{\alpha\alpha}(\textbf{q},\omega)$ at $T$=$0.003$ eV. Figure \ref{chiS} (a), (b), (c), and (d) show $\chi^{s}_{AA}(\textbf{q},0)$, $\chi^{s}_{BB}(\textbf{q},0)$, ${\rm Im}[\chi^{s}_{AA}(\textbf{q},\omega_{0})]$, and ${\rm Im}[\chi^{s}_{BB}(\textbf{q},\omega_{0})]$ in the $q_b$-$q_c$ plane, respectively. $\chi^{s}_{\alpha\alpha}(\textbf{q},0)$ is a real number. We fix $q_a$=$0$ in Fig. \ref{chiS} (a) and (c) and $q_a$=$0.2\pi$ in Fig. \ref{chiS} (b) and (d). Furthermore, $\omega_{0}$ is equal to $0.001$ eV. ${\rm Re}[\chi^{s}_{AA}(\textbf{q},0)]$ and ${\rm Im}[\chi^{s}_{AA}(\textbf{q},\omega_{0})]$ have very large values at $\textbf{q}$=$\textbf{0}$ because $\xi_{s}(\textbf{0})$=$0.999$. However, the BB component is difficult to be affected by $\xi_{s}(\textbf{0})$ but easily affected by $\xi_{s}(\textbf{Q})$.
$\textbf{Q}$ is the wavenumber at which $\chi^{0}_{BB}(\textbf{q},0)$ and $\chi^{s}_{BB}(\textbf{q},0)$ have the maximum values.
$\chi^{s}_{AA}(\textbf{0},0)$ and ${\rm Im}[\chi^{s}_{AA}(\textbf{0},\omega_{0})]$ are much larger than $\chi^{s}_{BB}(\textbf{Q},0)$ and ${\rm Im}[\chi^{s}_{BB}(\textbf{Q},\omega_{0})]$ because $\xi_{s}(\textbf{0})>\xi_{s}(\textbf{Q})$ and the spin susceptibility obtained using RPA is determined by $1/\left(1-\xi_{s}(\textbf{q})\right)$.
$\chi^{s}_{AA}(\textbf{0},0)$ and ${\rm Im}[\chi^{s}_{AA}(\textbf{0},\omega_{0})]$ in Fig. \ref{chiS} (a) and (c) decrease with temperature. On the other hand, $\chi^{s}_{BB}(\textbf{Q},0)$ and ${\rm Im}[\chi^{s}_{BB}(\textbf{Q},\omega_{0})]$ in Fig. \ref{chiS} (b) and (d) slowly increase and the peaks become broad with temperature.
These behavior result from temperature dependence of $\xi_s(\textbf{q})$ in Fig. \ref{Stoner}

\subsection{Knight shift and $1/T_{1}T$ in the presence of $U$}

In this subsection, we solve Eq. \ref{KS_formula} and \ref{T1T_formula} to investigate the effects of the fluctuations on the Knight shift and $1/T_{1}T$.

\begin{figure}[htpb]
\begin{center}
\includegraphics[width=60mm]{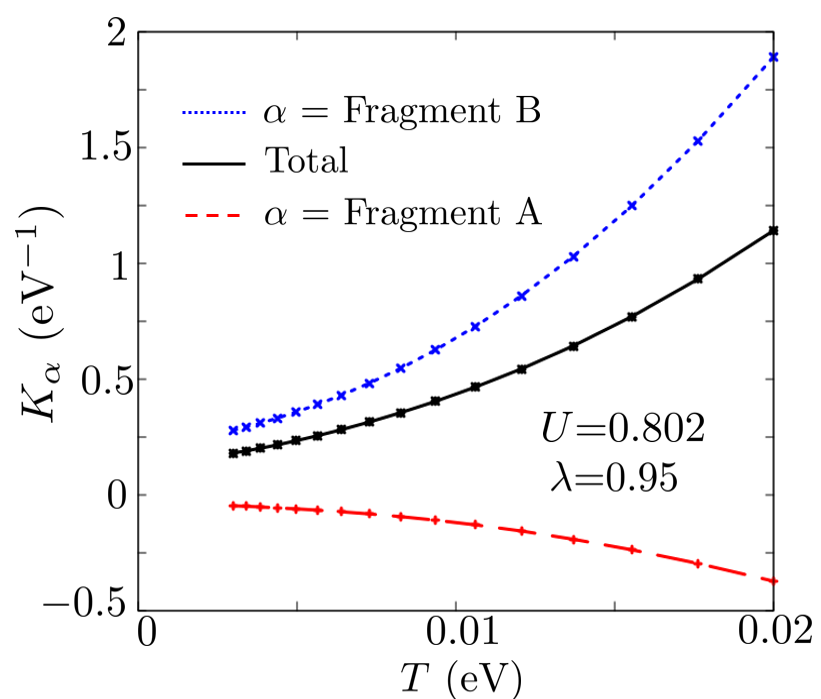}
\end{center}
\caption{ Temperature dependence of the Knight shift $K_{\alpha}$ for $U$=$0.802$ and $\lambda$=$0.95$. The red dashed and blue dotted lines show $K_{A}$ and $K_{B}$, respectively. The black solid line shows the total Knight shift $K_{\rm tot}$=$K_{A}+K_{B}+K_{C}$.}
\label{KSS}
\end{figure}

Fig. \ref{KSS} shows the temperature dependence of the Knight shift, where $U$=$0.802$ and $\lambda$=$0.95$. $K_{B}$ in the presence of $U$ is larger than that in the absence of $U$. However, $K_{A}$ and $K_{C}$ in the presence of $U$ are smaller than those in the absence of $U$. In the case of $U$=$0.802$ and $\lambda$=$0.95$, $K_{A}$ and $K_{C}$ are negative, while $K_{B}$ and $K_{tot}(=K_{A}+K_{B}+K_{C})$ are positive. Similar behavior was previously observed in the organic conductor $\alpha$-(BEDT-TTF)$_{2}$I$_{3}$.\cite{A.Kobayashi2013}

$K_{A}$ and $K_{C}$ don't increase at low temperature, although the Stoner factor $\xi_{s}(\textbf{0})$ is almost $1$ at $T$=$0.003$ eV. The behavior of the Knight shift is understood by considering the off-diagonal elements of $\hat{\chi}^{s}(\textbf{0},0)$.
Because $\chi^{s}_{AA}(\textbf{0},0)$ and $\chi^{s}_{AC}(\textbf{0},0)$ have opposite signs in Fig. \ref{chiS_00}(a), their cancelation prevents the increase of the Knight shift in Eq. \ref{KS_formula}.
In other words, the $\textbf{q}$=$\textbf{0}$ spin fluctuations are not observed in the Knight shift because of the intra-molecular antiferromagnetic fluctuations.

\begin{figure}[htpb]
\begin{center}
\includegraphics[width=60mm]{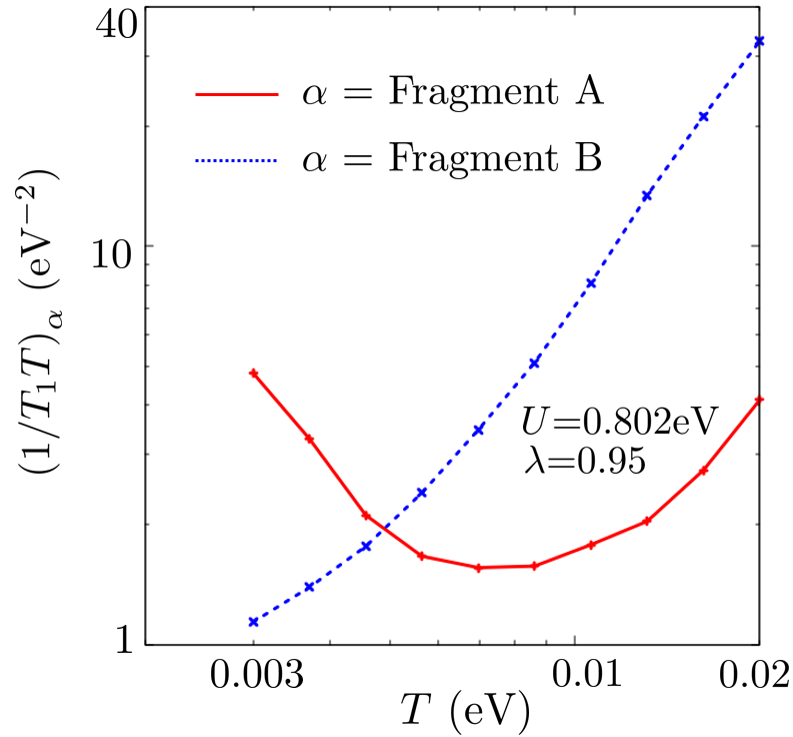}
\end{center}
\caption{ Temperature dependence of $(1/T_{1}T)_{\alpha}$ for $U$=$0.802$ and $\lambda$=$0.95$. The red solid and blue dotted lines show $(1/T_{1}T)_{A}$ and $(1/T_{1}T)_{B}$, respectively. }
\label{T1TS}
\end{figure}

Next, we solve Eq. \ref{T1T_formula}. The spin-lattice relaxation rate, $1/T_{1}T$, is determined by $\sum_{\textbf{q}}{\rm Im}[\chi^{s}_{\alpha\alpha}(\textbf{q},\omega_{0})]$. ${\rm Im}[\chi^{s}_{AA}(\textbf{q},\omega_{0})]$ and ${\rm Im}[\chi^{s}_{BB}(\textbf{q},\omega_{0})]$ are shown in Fig. \ref{chiS}.
Figure \ref{T1TS} shows the temperature dependence of $1/T_{1}T$, where $\lambda$=$0.95$ and $U$=$0.802$.
At high temperature, $1/T_{1}T$ values for all orbitals decrease with decreasing $T$. However, at low temperature, the $(1/T_{1}T)_{A}$ starts to increase because orbital A is easily affected by $\xi_{s}(\textbf{0})$. The behavior of $(1/T_{1}T)_{C}$ is identical to that of $(1/T_{1}T)_{A}$ because of space-inversion symmetry.
Although $(1/T_{1}T)_{B}$ is difficult to be affected by $\xi_{s}(\textbf{0})$ in Fig. \ref{T1TS}, $\xi_{s}(\textbf{0}) \rightarrow 1$ makes $(1/T_{1}T)_{B} \rightarrow \infty$.
This is because $\xi_{s}(\textbf{0}) \rightarrow 1$ makes $1/\left(1-\xi_{s}(\textbf{0}) \right) \rightarrow \infty$ and RPA imposes the factor $1/\left(1-\xi_{s}(\textbf{0}) \right)$ on the spin susceptibility (Eq. \ref{xi_and_chiS}).
For $T$$\gtrsim$$0.005$ eV, $(1/T_{1}T)_{B}$ is more dominant than $(1/T_{1}T)_{A}$ and $(1/T_{1}T)_{C}$ because $\xi_{s}(\textbf{Q})$ slowly increases with $T$ and orbital B is easier to be affected by $\xi_{s}(\textbf{Q})$.
$\textbf{Q}$ is the wavenumber at which $\chi^{0}_{BB}(\textbf{q},0)$ and $\chi^{s}_{BB}(\textbf{q},0)$ have the maximum values.

In the case of a small $\lambda$, such as $\lambda$=$0.79$, $\xi_{s}(\textbf{Q})$ is larger than $\xi_{s}(\textbf{0})$, and SDW can be induced. However, the incommensurate spin fluctuations in this material are suppressed at low temperature in Fig. \ref{Stoner}. Similar behavior occurs in the case of a small $\lambda$. If $U$ is sufficiently large, $\xi_{s}(\textbf{Q})$ reaches $1$ at low temperature. However, the magnetic transition to the SDW phase would have already been induced at a high temperature. Thus, it is difficult to explain the upturn of the $1/T_{1}T$ curve near $30$ K, which is observed in the $^{13}$C-NMR experiment, by the incommensurate spin fluctuations.
We calculate $\lambda$ using the \textsc{respack} program.
Because $\lambda$=$0.79$ is obtained as the ratio of diagonal elements with the unscreened on-site Coulomb interaction while $\lambda$=$0.95$ is obtained as that with the screened on-site Coulomb interaction, we consider that $\lambda$=$0.95$ is the more realistic ratio.

\begin{figure}[htpb]
\begin{center}
\includegraphics[width=85mm]{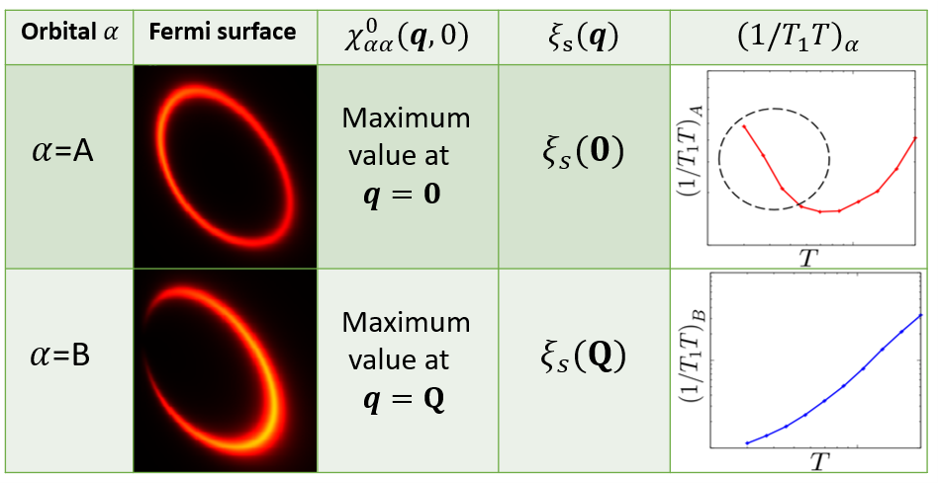}
\end{center}
\caption{ Correspondence table between the fragment orbitals, shape of the Fermi surface, momenta at which the non-interacting spin susceptibilities have peaks, Stoner factors, and $1/T_1T$.
}
\label{TB2}
\end{figure}

In this subsection, we showed that the Knight shift does not increase at low temperature because of the intra-molecular antiferromagnetic fluctuations. We also showed that the $(1/T_{1}T)_{A}$ and $(1/T_{1}T)_{C}$ start to increase at low temperature because of the behavior of $\xi_{s}(\textbf{0})$. They are dominant in $T$$\lesssim$$0.005$ eV, whereas $(1/T_{1}T)_{B}$ is dominant in $T$$\gtrsim$$0.005$ eV because of the temperature dependence of $\xi_{s}(\textbf{Q})$ and fragment-orbital-dependence of spin susceptibilities. Fig. \ref{TB2} summarizes the fragment-orbital dependence of the Fermi surface, non-interacting spin susceptibilities, Stoner factors, and $1/T_{1}T$. Fig. \ref{TB2} shows the important factors for orbitals A and B. $\xi_{s}(\textbf{0})$ is the main contributor to orbital A and causes the upturn of the $(1/T_{1}T)_{A}$ curve at low temperature. However, the contribution of $\xi_{s}(\textbf{0})$ to orbital B is small. Orbital B is sensitive to $\xi_{s}(\textbf{Q})$, which dominantly contributes to $(1/T_{1}T)_{B}$ at high temperature but does not contribute to orbital A as much.
These fragment-orbital-dependent magnetic properties are caused by the presence of ZR because ZR biases $\rho_{B}(\textbf{k},0)$ in Eq. \ref{spectrum} to a part of the Fermi surface.

\section{conclusion}
In this study, we found that multiple fragment orbitals play important roles in the magnetic properties of [Ni(dmdt)$_2$]. On orbitals A and C, which are unevenly distributed toward one side of a molecule, the $\textbf{q}$=$\textbf{0}$ spin fluctuations are enhanced, whereas the incommensurate spin fluctuations are enhanced on orbital B, which is centered on the Ni atom. Because of the $\textbf{q}$=$\textbf{0}$ spin fluctuations, the A and C components of $1/T_{1}T$ start to increase as $T$ decreases at low temperature.
However, the $\textbf{q}$=$\textbf{0}$ spin fluctuations do not affect the Knight shift, because they are intra-molecular antiferromagnetic fluctuations.
The reason why the \textbf{q}=\textbf{0} spin fluctuations is enhanced is understood from the perturbation process and the off-diagonal elements of the non-interacting spin susceptibility. The incommensurate spin fluctuations dominantly contribute to the B component of $1/T_{1}T$ at high temperature. These fragment-orbital-dependent quantities result from the presence of ZR.
If no ZR exists in the Brillouin zone, the BB components of the spin susceptibilities do not have the maximum value at the incommensurate wavenumber, because the spectral weight of fragment orbital B at $E_{F}$ may be similar to those of A and C.
Because ZR biases $\rho_{B}(\textbf{k},0)$ in Eq. \ref{spectrum} to a part of the Fermi surface, the BB component of the spin susceptibility has a maximum value at the incommensurate wavenumber $\textbf{q}=\textbf{Q}$. Thus, the wavenumber dependence of the spin susceptibilities is different between the BB component and AA(CC) component.
ZR is a characteristic of materials with a Dirac nodal line system described by an $n$-band model($n\geq 3$). Thus, it is expected that the fragment-orbital-dependent properties due to ZR will be found in other Dirac nodal line systems. Moreover, it is predicted that transition-metal substitution in the Ni(dmdt)$_2$ molecule controls spin fluctuations because it changes $\lambda$ and $U$.
In the two-dimensional Dirac electron system under the charge-neutral condition, the spin fluctuations are weak because the Fermi surface is identical to the Dirac points.
On the other hand, the spin fluctuations are enhanced in [Ni(dmdt)$_2$] by the Fermi surface. The Fermi surface arises from transfer integrals in the nodal line direction. This is the three-dimensionality of this material.

In the $^{13}$C-NMR experiment, $1/T_{1}T$ has a peak at $T$$\sim$$30$ K. The experiment was performed for a sample in which C atoms were replaced with $^{13}$C. Fig. \ref{fragment}(b) shows a Ni(dmdt)$_2$ molecule, and the red dashed circles surround $^{13}$C atoms. Therefore, orbitals A and C mainly contribute to the physical quantities observed in the $^{13}$C-NMR experiment. Thus, our calculation of $1/T_{1}T$ is almost consistent with the experiment. Furthermore, we expect that the B component of $1/T_{1}T$ can be observed by experiments using a sample in which $^{12}$C atoms near the Ni atom are substituted by $^{13}$C. On the other hand, the A and C components of the Knight shift obtained in our calculation are negative, but the Knight shift observed in the $^{13}$C-NMR experiment is positive. Therefore, we consider the following possible electronic states at $T$$\lesssim$$30$ K.
The first is the $\textbf{q}$=$\textbf{0}$ magnetic ordered state, which is the intra-molecular antiferromagnetism is realized at $T$$\lesssim$$30$ K. In this case, it is considered that the Knight shift observed in the experiment is attributed to the sum of $K_{A}$ and $K_{B}$, which is positive.
In the second possible electronic state, $U$ is not so large that $K_{A}$ is negative. In this case, it is considered that another ordered state is induced at $T$$\lesssim$$30$ K and that the $1/T_{1}T$ curve is upturned by the fluctuations corresponding to the order. Examples of such orders are bond order and topological order.

\begin{acknowledgments}
The authors thank T. Sekine, T. Hatamura, K. Sunami, K. Miyagawa, K. Kanoda, B. Zhou, Akiko Kobayashi, and K. Yoshimi for fruitful discussions.
This work was financially supported by MEXT (JP) JSPS (Grant No. 15K05166) and JST SPRING (Grant No. JPMJSP2125). T. Kawamura acknowledges support from the ``Interdisciplinary Frontier Next-Generation Researcher Program of the Tokai Higher Education and Research System.'' The computation in this work was performed using the facilities of the Supercomputer Center, Institute for Solid State Physics, University of Tokyo.
\end{acknowledgments}

\end{document}